\documentclass[a4paper,11pt]{article}
\pdfoutput=1 
\usepackage{jheppub} 
\usepackage[T1]{fontenc} 
\usepackage{amssymb}
\usepackage{graphicx}
\usepackage{color}
\usepackage{tabls}
\usepackage{hyperref}
\usepackage{bm}
\usepackage{orcidlink}
\DeclareMathOperator{\Tr}{\mathrm{Tr}}

\DeclareMathOperator{\Z}{\mathbb{Z}}

\DeclareMathOperator{\dd}{{\rm{d}}}

\newcommand{\kB}{k_{\mbox{\tiny{B}}}}
\newcommand{\lambdai}{\lambda_{\mbox{\tiny{i}}}}
\newcommand{\lambdaf}{\lambda_{\mbox{\tiny{f}}}}
\newcommand{\ti}{t_{\mbox{\tiny{i}}}}
\newcommand{\tf}{t_{\mbox{\tiny{f}}}}
\newcommand{\Zi}{Z_{\mbox{\tiny{i}}}}
\newcommand{\Zf}{Z_{\mbox{\tiny{f}}}}
\newcommand{\Hi}{H_{\mbox{\tiny{i}}}}
\newcommand{\Hf}{H_{\mbox{\tiny{f}}}}
\newcommand{\Si}{S_{\mbox{\tiny{i}}}}
\newcommand{\Sf}{S_{\mbox{\tiny{f}}}}
\newcommand{\Ntherm}{N_{\mbox{\tiny{therm}}}}
\newcommand{\eq}[1]{\begin{align}\label{#1}}
\newcommand{\en}{\end{align}}
\newcommand{\eqar}[1]{\begin{align}\label{#1}}
\newcommand{\enar}{\end{align}}

\usepackage[english]{babel}
\usepackage{lmodern}

\usepackage{xcolor}

\usepackage{dsfont}
\usepackage{amsmath}
\usepackage{amsthm}
\usepackage{mathtools}
\usepackage{mathrsfs}
\usepackage{slashed}
\usepackage{physics}
\usepackage{tikz}
\usetikzlibrary{decorations.pathreplacing}
\usetikzlibrary{arrows.meta}

\usepackage{afterpage}

\usepackage{caption}
\usepackage{subcaption}
\usepackage{float}
\usepackage[format=plain,
            font=it]{caption}
\usepackage{multirow}

\newcommand{\densmat}[1]{\ket{#1}\bra{#1}}

\title{\boldmath Entanglement entropy from non-equilibrium Monte~Carlo simulations}

\author[a]{Andrea~Bulgarelli,\orcidlink{0009-0002-2917-6125}}
\author[a]{and Marco~Panero\orcidlink{0000-0001-9477-3749}}

\affiliation[a]{Physics Department, University of Turin \& INFN, Turin unit\\Via Pietro Giuria 1, I-10125 Turin, Italy}

\emailAdd{andrea.bulgarelli@unito.it}
\emailAdd{marco.panero@unito.it}

\abstract{We study the entanglement entropy in lattice field theory using a simulation algorithm based on Jarzynski's theorem. We focus on the entropic c-function for the Ising model in two and in three dimensions: after validating our algorithm against known analytical results from conformal field theory in two dimensions, we present novel results for the three-dimensional case. We show that our algorithm, which is highly parallelized on graphics processing units, allows one to precisely determine the subleading corrections to the area law, which have been investigated in many recent works. Possible generalizations of this study to other strongly coupled theories are discussed.}

\begin{document} 


\maketitle
\flushbottom

\section{Introduction}
\label{sec:introduction}

Entanglement is a defining characteristic of quantum systems, and has far-reaching implications that range from low- to high-energy physics. On the one hand, condensed-matter systems close to a quantum critical point are expected to be characterized by a strongly entangled ground state~\cite{Osborne:2002zz, Vidal:2002rm, Latorre:2003kg, Kitaev:2005dm, Amico:2007ag, Laflorencie:2015eck}. On the other hand, in strongly coupled gauge theories (such as quantum chromodynamics, the theory of the strong interaction in the Standard Model of elementary particle physics), entanglement is expected to reveal whether the theory is in a deconfined or in a confining phase~\cite{Nishioka:2006gr, Klebanov:2007ws, Liu:2012eea, Kol:2014nqa} (or whether it has a finite mass gap~\cite{Jokela:2020wgs}). In the context of the physics of black holes, the Hawking radiation~\cite{Hawking:1975vcx} is entangled with degrees of freedom within the black hole, so entanglement is directly relevant for the information loss paradox~\cite{Mathur:2009hf, Unruh:2017uaw, Raju:2020smc}. In quantum gravity, it was argued that the entanglement even leads to the emergence of classically connected spacetimes~\cite{VanRaamsdonk:2010pw}.

In addition, entanglement also plays an essential r\^ole in quantum information science~\cite{nielsen2000quantum}; in particular, being intimately related to the speed-up in information processing and communication that quantum computers can attain with respect to classical ones (i.e., to quantum advantage), it has disruptive technological potential.

While defining a quantitative measure of the entanglement of a multipartite state is not a completely trivial matter~\cite{Horodecki:2009zz}, the general requirements it has to satisfy (including, in particular, monotonicity under local transformations of the state) are known~\cite{Vidal:1998re}; for bipartite systems, the entanglement can be quantified in terms of the entanglement entropy~\cite{Bennett:1995tk}, which is the von~Neumann entropy of the reduced density matrix associated with either of the subsystems that make the system. In turn, the von~Neumann entropy can be obtained as a limit of the R\'enyi entropy: given a quantum system with density matrix $\rho$, if one considers a bipartition of the system in two parts, $A$ and $B$, the R\'enyi entropy of order $n$ of the subsystem $A$ is defined as 
\begin{align}
S_n(A) = \frac{1}{1-n}\log\Tr\rho_A^{n},
\end{align}
where $\rho_A = \Tr_B \rho$ is the reduced density matrix for $A$. In the limit $n \to 1$ the R\'enyi entropy tends to the von~Neumann entropy for the subsystem $A$,
\begin{align}
S(A) = -\Tr(\rho_A\log\rho_A).
\end{align}
An interesting feature of the entanglement entropy is that it satisfies an ``area law''~\cite{Eisert:2008ur}: $S_n$ is proportional to the area of the boundary $\partial A$ separating the subsystems $A$ and $B$, which is usually called the entangling surface. This property is particularly intriguing, as it parallels an analogous property of the entropy of black holes~\cite{Bekenstein:1973mi,Hawking:1975vcx}.

The area term, however, is ultraviolet-divergent, hence a proper computation of the entanglement entropy requires a suitable regularization and renormalization procedure (which, in general, depends on the shape of the region $A$) and the physical information about the system is contained in the subleading contributions to the entropy. For a system defined in $D = d+1$ spacetime dimensions, if one considers a geometry in which the entangling surface is independent from the size of the subsystem $A$, for example if $A$ is a slab with length $l$ in the direction separating $A$ and $B$, and maximally extended in the other directions, the derivative of the entanglement entropy with respect to $l$ is not ultraviolet-divergent. Accordingly, one can define the entropic (R\'enyi) c-functions as~\cite{Casini:2004bw, Casini:2006es, Nishioka:2006gr}
\begin{align}
C_n(l) = \frac{l^{D-1}}{|\partial A|}\frac{\partial S_n}{\partial l},
\label{definition_entropic_c-function}
\end{align}
where $|\partial A|$ is the area of the entangling surface. The entropic c-function encodes the physical information contained in the R\'enyi entropies and it has been proven that, in $D = 2$, $3$, $4$ and for specific geometries, it is monotonically decreasing along the renormalization group trajectories~\cite{Casini:2004bw, Casini:2012ei, Casini:2017vbe}, hence it provides a suitable measure of the number of degrees of freedom in the theory~\cite{Zamolodchikov:1986gt}; for a discussion about the connection of these quantities with the irreversibility of the renormalization group flows in quantum field theories with planar defects, see also the recent ref.~\cite{Casini:2023kyj}.

The computation of entanglement entropy is a challenging task: analytically, it can be carried out only for systems with a large amount of symmetry, such as conformal field theories (CFT) in low-dimensional systems~\cite{Calabrese:2004eu, Cardy:2007mb, Calabrese:2009qy, Callebaut:2023fnf}, or using the conjectured gauge/gravity correspondence~\cite{Maldacena:1997re, Gubser:1998bc, Witten:1998qj}, whereby the entanglement entropy is obtained from the area of minimal surfaces in an anti-de~Sitter (AdS) spacetime~\cite{Ryu:2006bv, Ryu:2006ef}. The entanglement entropy has also been studied numerically, using different methods and in different models (an incomplete list of relevant works includes refs.~\cite{Buividovich:2008kq, Buividovich:2008gq, Buividovich:2008yv, Caraglio:2008pk, Alba:2009ek, Gliozzi:2009zc, Nakagawa:2009jk, Nakagawa:2010kjk, Hastings:2010zka, Alba:2011fu, Humeniuk:2012xg, Grover:2013nva, Coser:2013qda, Drut:2015aoa, Itou:2015cyu, Alba:2016bcp, Rabenstein:2018bri, DEmidio:2019usm, Zhao:2021njg, Zhao:2021ghz, Rindlisbacher:2022bhe, DaLiao:2023pdn, Bringewatt:2023xxc}), but also in this case one has to cope with non-trivial difficulties; part of the problems stem from non-local nature of the observable, from the fact that several of these numerical techniques are hampered by limited scalability to systems in more than one spatial dimension, and from the prohibitively large amount of computational time that is necessary, to properly sample the space of configurations.

In recent years, our group has published a series of high-precision lattice calculations of different physical quantities~\cite{Caselle:2016wsw, Caselle:2018kap, Francesconi:2020fgi, Caselle:2022acb} using non-equilibrium Monte~Carlo simulations based on Jarzynski's theorem~\cite{Jarzynski:1996oqb, Jarzynski:1997ef}. The latter is one of the exact equalities that, during the last decade of the past century, highlighted the connections between deviations from thermodynamic equilibrium and entropy production~\cite{Evans:1993po, Evans:1993em, Gallavotti:1994de, Gallavotti:1995de, Crooks:1997ne, Crooks:1999ep} (see also refs.~\cite{Ritort:2004wf, MariniBettoloMarconi:2008fd}). Jarzynski's theorem has many remarkable implications, including ones that are relevant for quantum information and quantum chaos~\cite{Esposito:2009zz, DAlessio:2016rwt, Campisi:2016qlj, Halpern:2016zcm, Halpern:2017abm, Mori:2018qjb, Chenu:2018spm}. For its applications in the context of numerical simulations, it has been recently stressed~\cite{Caselle:2022acb} that Jarzynski's theorem has a direct connection with machine learning approaches (specifically, with a general class of generative models called stochastic normalizing flows~\cite{Wu:2020snf}) as well as with annealed importance sampling~\cite{Neal:1998ais}. It is becoming increasingly clear that non-equilibrium Monte Carlo calculations based on this theorem can be a reliable tool for studies of the entanglement entropy, too~\cite{Alba:2016bcp, DEmidio:2019usm, Zhao:2021njg, DaLiao:2023pdn}.

In this work we present a new algorithm, combining Jarzynski's theorem and the replica trick~\cite{Calabrese:2004eu}, for direct Monte~Carlo calculations of the entropic c-function on the lattice. Our algorithm is implemented in the CUDA code and is run on machines equipped with graphics processing units (GPUs). Being based on Monte~Carlo simulations, our algorithm works in any integer dimension, and we discuss its application to the Ising model~\cite{Ising:1925em}. In general, the latter is defined in terms of a collection of $\sigma_i$ variables on the sites $i$ of an arbitrary lattice and taking values $\sigma_i=\pm 1$; in our notations (assuming natural units $\hbar=c=\kB=1$), the reduced Hamiltonian of a configuration is
\begin{align}
\label{Ising_model}
\frac{H(\sigma)}{T}=-\beta\sum_{\langle i, j\rangle} \sigma_i \sigma_j,
\end{align}
where $T$ denotes the temperature, and the sum runs over the nearest-neighbor sites on the lattice. For simplicity, here we neglect the possibility of a coupling to an external field, so that the model has a global $\Z_2$ symmetry, as the Hamiltonian is invariant under a transformation in which all spins are flipped. We consider an isotropic square lattice in $D=2$ (testing our algorithm against analytical results) and an isotropic cubic lattice in $D=3$. In the latter case we show that with our algorithm it is possible to efficiently extract the subleading terms beyond the area law, that have been the subject of intense studies in recent years~\cite{Fradkin:2006mb, Hsu:2008af, Ju:2012esi, Stephan:2009sae, Stephan:2010reo, Stephan:2013eig, Inglis:2013eaa, Chen:2014zea, Sahoo:2015hma, Chen:2016kjp, Kulchytskyy:2019hft}. In particular, we study how the entropic c-function at the critical point of the model varies, as a function of the size of the subsystem $A$. In addition, we also study the entropic c-function away from the critical point, investigating its dependence on the coupling of the model.

The structure of this article is the following. In section~\ref{sec:replica_trick} we briefly review the replica trick and the analytical results that can be obtained for conformal field theories in two dimensions, while in section~\ref{sec:protocol} we describe our algorithm. The results for the Ising model in two and three dimensions are presented in section~\ref{sec:Ising_2D} and in section~\ref{sec:Ising_3D}, respectively. Section~\ref{sec:conclusions} is devoted to a summary of our findings and to some comments about future generalizations of this work, in particular for strongly coupled lattice gauge theories. Finally, the appendix contains a proof of Jarzynski's theorem (section~\ref{app:jarzynski_s_theorem}), the discussion of a different type of simulation strategy for the three-dimensional model (section~\ref{app:a_different_protocol_for_the_three-dimensional_case}), and a quantitative analysis of the systematic uncertainties associated with the derivative discretization in our algorithm (section~\ref{app:systematic_uncertainty_from_the_derivative_discretization}).

\section{Replica trick and exact results}
\label{sec:replica_trick}

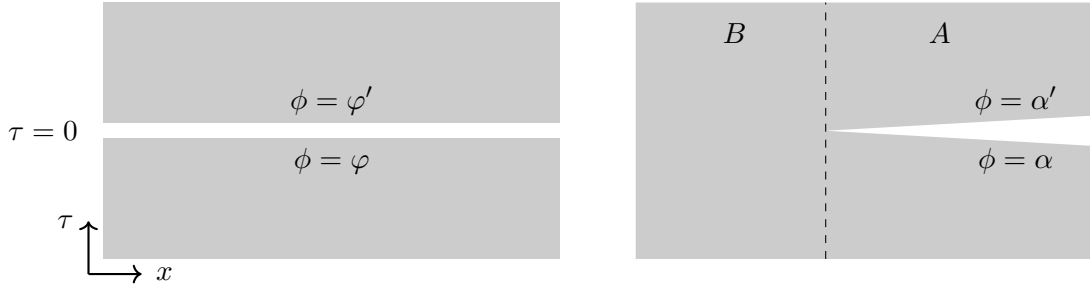
\begin{figure}
\centering

\begin{tikzpicture}

\fill[gray!40!white] (-1,1.8)rectangle(5,3.4);
\draw (2,2.1)node{$\phi=\varphi'$};
\fill[gray!40!white] (-1,0)rectangle(5,1.6);
\draw (2,1.3)node{$\phi=\varphi$};
\draw (-1.8,1.7) node{$\tau = 0$};
\draw[->, thick] (-1.2,-.2)--(-1.2,.5);
\draw[->, thick] (-1.2,-.2)--(-.5,-.2);
\draw (-1.5,.5) node{$\tau$};
\draw (-.2,-.2) node{$x$};
\fill[gray!40!white] (6,0)--(12,0)--(12,1.5)--(8.5,1.7)--(12,1.9)--(12,3.4)--(6,3.4)--(6,0);
\draw (11,2.1)node{$\phi=\alpha'$};
\draw (11,1.3)node{$\phi=\alpha$};
\draw[dashed] (8.5,0)--(8.5,3.4);
\draw (7.3,3) node{$B$};
\draw (10,3) node{$A$};

\end{tikzpicture}

\caption{Left-hand-side panel: schematic representation of the matrix element of the density matrix between two field configurations. Right-hand-side panel: a representation of the matrix element of the reduced density matrix. Gray areas correspond to regions of spacetime where field configurations are integrated upon.}
\label{fig:cut_plane_with_field_configurations}
\end{figure}

Consider the ground state $\ket{0}$ of a generic quantum field theory and the related density matrix $\rho = \densmat{0}$. Our first aim is to express a generic matrix element $\mel{\varphi ' }{\rho}{\varphi} = \braket{\varphi '}{0}\braket{0}{\varphi}$ in the path-integral formalism. If one considers two generic field configurations $\varphi(\mathbf{x},\tau=0)$ and $\varphi(\mathbf{x},\tau=-T)$ one can write
\begin{align}
\braket{\varphi(\mathbf{x},0)}{\varphi'(\mathbf{x},-T)} = %
\frac{1}{\sqrt{Z}} \int_{\phi(\mathbf{x},-T) = \varphi(\mathbf{x},-T)} ^ {\phi(\mathbf{x},0) = \varphi(\mathbf{x},0)} D\phi \: e^{-S[\phi]}
\end{align}
where $Z$ is a normalization constant. In the $T\to \infty$ limit, the contribution from excited states becomes negligible, hence
\begin{align}
\braket{\varphi(\mathbf{x},0)}{0} = \frac{1}{\sqrt{Z}} \int_{\phi(\mathbf{x},-\infty) = 0} ^ {\phi(\mathbf{x},0) = \varphi(\mathbf{x},0)} D\phi \: e^{-S[\phi]},
\end{align}
therefore the matrix element of the density matrix reads
\begin{align}
\mel{\varphi'}{\rho}{\varphi} = \frac{1}{Z} \int_{\phi(\mathbf{x},-\infty) = 0} ^ {\phi(\mathbf{x},0^-) = \varphi(\mathbf{x},0^-)} \int_{\phi(\mathbf{x},0^+) = \varphi(\mathbf{x},0^+)}^{\phi(\mathbf{x},\infty) = 0} D\phi \: e^{-S[\phi]}
\end{align}
The previous expression can be schematically depicted as in fig.~\ref{fig:cut_plane_with_field_configurations}: one is integrating over all spacetime, leaving open a cut at $\tau = 0$ where the field configurations are fixed. From the condition $\Tr\rho = 1$, one can fix the normalization factor to be $Z = \int D\phi e^{-S[\phi]}$, i.e., the partition function.

Now let us introduce a partition of the system into two subsystems such that
\begin{align}
\ket{\varphi} = \ket{\alpha}_A \oplus \ket{\beta}_B .
\end{align}
The reduced density matrix reads
\begin{align}
\mel{\alpha'}{\rho}{\alpha} = \frac{1}{Z}\int D\beta D\beta' \mel{\alpha'\oplus\beta'}{\rho}{\alpha\oplus\beta}\delta(\beta-\beta').
\end{align}
Tracing over the degrees of freedom of $B$ corresponds to identifying $\beta(\mathbf{x},0) = \beta'(\mathbf{x},0)$ and integrating over $\beta$; this can be interpreted as joining the edges of the cut in correspondence of the subsystem $B$.

From the previous expression, one can easily obtain the matrix element of the square of the reduced density matrix as
\begin{align}
\mel{\alpha'_2}{\rho^2}{\alpha_1} = \frac{1}{Z^2}\int D\alpha_2 D\alpha'_1 \mel{\alpha'_2}{\rho}{\alpha_2}\mel{\alpha'_1}{\rho}{\alpha_1}\delta(\alpha_2 - \alpha'_1).
\end{align}
Again this operation can be given a geometric interpretation: one introduces a second cut space (a replica), with field configurations $\alpha'_2$ and $\alpha_2$ at the upper and lower edges of the cut, respectively. Then the upper edge of the first replica is joined with the lower edge of the second one, by identifying $\alpha_2 = \alpha'_1$ and integrating over $\alpha_2$.

%
%
%
\noindent
%
%
%
It is now clear that a matrix element of the $n$-th power of the reduced density matrix can be expressed as the path integral over $n$ copies of the cut space, where the upper edge of the cut $i$ is identified with the lower edge of the cut $i+1$, for $i = 2,\dots,n-1$, while in the first and last edges the field configurations are held fixed. Finally, if one joins the remaining two edges, one obtains the expression
\begin{align}
\Tr\rho^n = \frac{Z_n}{Z^n},
\label{trace_of_rho_n_as_ratio_of_partition_functions}
\end{align}
where $Z_n$ is the partition function of the full replica space. Hence, the R\'enyi entropies can be written as
\begin{align}
S_n = \frac{1}{1-n}\log\frac{Z_n}{Z^n}.
\end{align}
Accordingly, the derivative of the R\'enyi entropy with respect to $l$ can be written as
\begin{align}
\frac{\partial S_n}{\partial l} = \frac{1}{n-1}\lim_{\epsilon\to 0}\frac{1}{\epsilon}\log\frac{Z_n(l)}{Z_n(l+\epsilon)}.
\label{definition_of_derivative_of_Renyi_entropy}
\end{align}
The replica trick is a powerful tool for analytical calculations: in two-dimensional conformal field theories it allows one to determine the scaling form of the R\'enyi entropies in a number of different geometries~\cite{Calabrese:2004eu, Calabrese:2009qy}. In a planar geometry, if the subsystem $A$ is a segment of length $l$, the R\'enyi entropies read
\begin{align}
S_n(l) = \frac{c}{6}\left( 1 + \frac{1}{n} \right)\log\left( \frac{l}{a} \right) + k_n
\label{Renyi_entropy_on_a_plane}
\end{align}
where $a$ is an ultraviolet regulator, $c$ the central charge of the theory and $k_n$ a constant (that depends on $n$); due to the presence of the logarithm, $k_n$ can be modified by rescaling $a$, hence its value is not universal.

Starting from eq.~(\ref{Renyi_entropy_on_a_plane}), one can conformally map the plane into a cylinder of spatial length $L$, with the cut orthogonal to the axis of the cylinder. This corresponds to a one-dimensional, finite-size quantum system at zero temperature, and the R\'enyi entropies become
\begin{align}
S_n(l,L) = \frac{c}{6}\left( 1 + \frac{1}{n} \right) \log(\frac{L}{\pi a}\sin(\frac{\pi l}{L})) + k_n.
\label{Renyi_entropy_on_a_cylinder}
\end{align}
Equation~(\ref{Renyi_entropy_on_a_cylinder}) can be obtained from eq.~(\ref{Renyi_entropy_on_a_plane}) by replacing the length of the cut with the arc-length, $l\to\frac{L}{\pi}\sin(\frac{\pi l}{L})$, hence it is invariant under $l\to L-l$, i.e., if we exchange $A$ and $B$. This is a consequence of a much more general property of R\'enyi entropies, namely, if $A \cup B$ is in a pure state then~\cite{Headrick:2019eth}
\begin{align}
S_n(A) = S_n(B).
\label{symmetry_of_Renyi_at_zero_temperature}
\end{align}
Equation~(\ref{Renyi_entropy_on_a_cylinder}) is an expression of fundamental importance for numerical studies, where only systems of finite size can be simulated.

We conclude this section with the expression of the entropic c-function for a $D=2$ CFT. In this case eq.~(\ref{definition_entropic_c-function}) becomes (we use $\partial A = 2$ since in one spatial dimension the entangling surface reduces to just two points)
\begin{align}
C_n(l) = \frac{l}{2}\frac{\partial S_n}{\partial l} = \frac{c}{12}\left( 1 + \frac{1}{n} \right),
\label{entropic_c-function_on_the_plane_2D}
\end{align}
hence the entropic c-function is a constant proportional to the central charge of the theory.

On the cylinder the presence of a multiplicative factor $l$ in the definition of the entropic c-function spoils the symmetry of the derivative of the R\'enyi entropy; in this work we use a symmetrized version of eq.~(\ref{definition_entropic_c-function}), trading $l$ for the arc-length, i.e.,
\begin{align}
C_n(l) = \left[\frac{L}{\pi}\sin(\frac{\pi l}{L})\right]^{D-1}\frac{1}{|\partial A|}\frac{\partial S_n}{\partial l}.
\label{symmetrised_entropic_c-function}
\end{align}
In the following, we will always refer to the entropic c-function as its symmetrized version. For a $D=2$ CFT one obtains
\begin{equation}
C_n(x) = \frac{c}{12}\left( 1 + \frac{1}{n}\right)\cos(\pi x),
\label{entropic_c-function_on_the_cylinder_2D}
\end{equation}
with $x = \frac{l}{L}$. The infinite-size result of eq.~(\ref{entropic_c-function_on_the_plane_2D}) is recovered at small $x$, with a leading correction of order $\order{x^2}$.

\section{Simulation algorithm for the entropic c-function}
\label{sec:protocol}
%
%
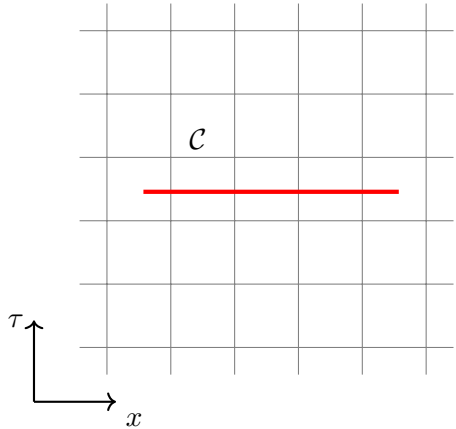
\begin{figure}[t]
\centering
\begin{tikzpicture}[scale=1.2]

\draw[step = .7cm, gray, thin] (-.2-.1, -.2-3+.1) grid (3.9-.1, 3.9-3+.1);
\draw[color=red, ultra thick](.4,-1.08)--(3.2,-1.08);
\draw[->, thick] (-.3 -.5,-3.1-.3)--(-.3-.5,-3.1-.3+.9);
\draw[->,thick] (-.3 -.5,-3.1-.3)--(-.3 -.5+.9,-3.1-.3);
\draw(-.3 -.5+1.1,-3.1-.5) node{$x$};
\draw(-.3 -.5-.2,-3.1-.5+1.1) node{$\tau$};
\draw(1,-.5) node{$\mathcal{C}$};

\end{tikzpicture}

\caption{A cut along the links of the dual lattice, where two different replicas are joined.}
\label{fig:replica_geometry_on_lattice}
\end{figure}\noindent
%
%
In this section we describe our Monte Carlo algorithm for the calculation of the entropic c-function. For the sake of clarity we will focus on the particular case of the Ising model, although the algorithm is easily generalizable to any spin model with nearest-neighbor interactions. Moreover, we start by considering a two-dimensional lattice. The extension to higher dimensions is straightforward and will be discussed at the end of the section.

Firstly, the lattice model has to be embedded in a replica geometry. Given $n$ copies of the lattice, we introduce a cut $\mathcal{C}$ in the dual lattice, lying along the spatial direction, as shown in fig.~\ref{fig:replica_geometry_on_lattice}. Then we introduce new links to connect spins on the upper side of the cut of one replica to the ones on the lower side of the cut of the following replica. The reduced Hamiltonian of the replica model thus reads
\begin{align}
\frac{H}{T} = - \sum_{k=1}^{n}\left\{\beta \sum_{\langle ij \rangle\not\perp\mathcal{C}}\sigma_i^{(k)}\sigma_j^{(k)} + \beta^{(k,k)}\sum_{\langle ij \rangle\perp \mathcal{C}}\sigma_i^{(k)}\sigma_j^{(k)} + \beta^{(k,k+1)}\sum_{\langle ij \rangle\perp\mathcal{C}}\sigma_i^{(k)}\sigma_j^{(k+1)}  \right\} .
\end{align}
We identify three different contributions:
\begin{itemize}
\item The first term takes into account interactions between spins in the same replica sheet connected through links that do not cross the cut.
\item The second term represents the coupling between spins in the same replica at opposite sides of the cut.
\item The last term represents couplings between spins in different replicas at opposite sides of the cut.
\end{itemize}
A decoupled system corresponds to a situation in which $\beta^{(k,k)}=\beta$ and $\beta^{(k,k+1)}=0$ for all $k$, while a fully coupled one corresponds to $\beta^{(k,k)}=0$ and $\beta^{(k,k+1)}=\beta$ for all $k$.

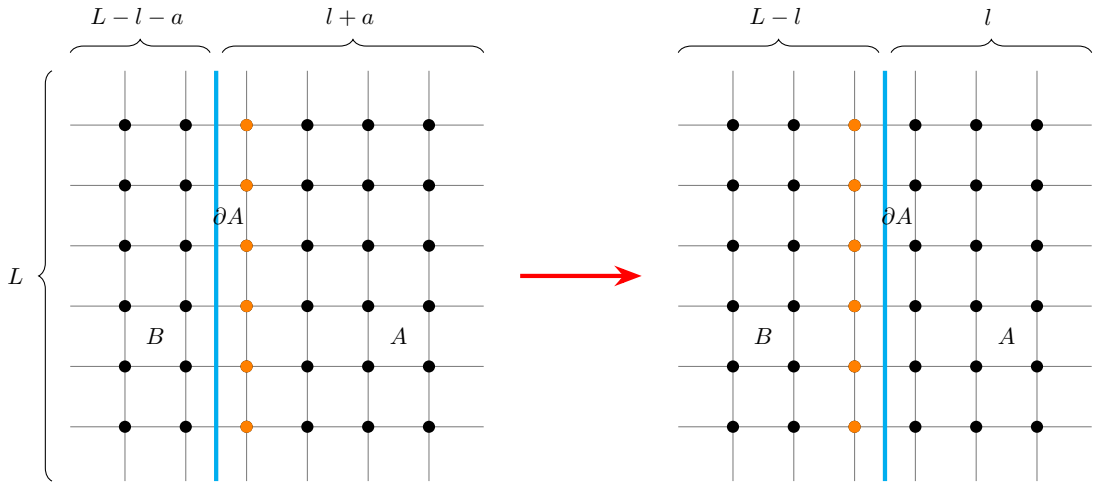
\begin{figure}
\centering
\vspace{0.5 cm}
\begin{tikzpicture}[scale=0.8, every node/.style={scale=0.8}]
\draw[step = 1cm, gray, thin] (-.9+ 10, -.9) grid (5.9+ 10, 5.9);
\foreach \x in {10,...,15}{
  \foreach \y in {0,...,5}{
    \fill[color=black] (\x, \y) circle (.1);
  }
}
\draw[cyan, ultra thick] (2.5+ 10,-.9)--(2.5+ 10,5.9);
\draw (4.5+10, 1.5)node{$A$};
\draw (2.7+10, 3.5)node{$\partial A$};
\draw (0.5+10, 1.5)node{$B$};

\foreach \y in {0,...,5}{
  \fill[color=orange] (2+10, \y) circle (.1);
}

\draw[decorate,decoration={brace,amplitude=5}](-.9,6.2)--(1.4,6.2);
\draw[decorate,decoration={brace,amplitude=5}](1.6,6.2)--(5.9,6.2);
\draw[decorate,decoration={brace,amplitude=5}](-1.2,-.9)--(-1.2,5.9);
\draw (.2,6.8)node{$L-l-a$};
\draw (3.7,6.8)node{$l+a$};
\draw (-1.8,2.5)node{$L$};
\draw[ultra thick,-Stealth, red] (6.5,2.5)--(8.5,2.5);
\draw[step = 1cm, gray, thin] (-.9, -.9) grid (5.9, 5.9);
\foreach \x in {0,...,5}{
  \foreach \y in {0,...,5}{
    \fill[color=black] (\x, \y) circle (.1);
  }
}
\draw[cyan, ultra thick] (1.5,-.9)--(1.5,5.9);
\draw (4.5, 1.5)node{$A$};
\draw (1.7, 3.5)node{$\partial A$};
\draw (0.5, 1.5)node{$B$};

\foreach \y in {0,...,5}{
  \fill[color=orange] (2, \y) circle (.1);
}

\draw[decorate,decoration={brace,amplitude=5}](-.9+10,6.2)--(2.4+10,6.2);
\draw[decorate,decoration={brace,amplitude=5}](2.6+10,6.2)--(5.9+10,6.2);
\draw (.7+10,6.8)node{$L-l$};
\draw (4.2+10,6.8)node{$l$};
\end{tikzpicture}
\caption{Sketch of our algorithm to compute the entropic c-function. The figure shows a spatial slice of a single replica; sites belonging to $A$ are connected to another replica through links in the Euclidean-time direction, while those belonging to $B$ are connected to the same replica. The ratio of partition functions in eq.~(\ref{entropic_c-function_in_code}) is computed from the exponential average of the work done on the system by varying the couplings of yellow sites from the initial geometry, where they are connected to another replica, to the one at the end of the non-equilibrium evolution, where they are connected to the same replica.}
\label{fig:higher_dimensions_c-function_algorithm}
\end{figure}

Before describing our algorithm for calculations of the entropic c-function, it is useful to briefly review the algorithm proposed in ref.~\cite{Alba:2016bcp} for the calculation of R\'enyi entropies, which is also based on Jarzynski's equality. In this case the initial partition function $\Zi$ corresponds to the partition function of $n$ decoupled replicas, while $\Zf$ is the partition function of a system of fully coupled replicas through a cut of length $l$ lattice spacings. The idea is to drive the system out of equilibrium by varying the couplings $\beta^{(k,k)}$ from $\beta$ to $0$ and, at the same time, the couplings $\beta^{(k,k+1)}$ from $0$ to $\beta$. More specifically the couplings of the model can be written as
\begin{align}
\beta^{(k,k')} = \begin{cases}
& \beta \delta_{k,k'} \qquad\qquad\qquad\qquad\qquad\qquad\:\: t < \ti\\
& \beta\frac{t-\tf}{\tf - \ti} (\delta_{k+1,k'} - \delta_{k,k'}) + \beta\delta_{k,k'} \qquad t \geq \ti
\label{off-equilibrium_evolution_of_the_couplings}
\end{cases}
\end{align}
and the work performed on the system in one Monte Carlo step reads
\begin{align}
\delta W = \frac{1}{\tf - \ti} \sum_{k,\langle ij \rangle \perp \mathcal{C}} \left\{ \sigma_i^{(k)}\sigma_j^{(k+1)} - \sigma_i^{(k)}\sigma_j^{(k)}  \right\}.
\end{align}
In contrast, the idea of our algorithm is instead to calculate the derivative of the R\'enyi entropy in eq.~\eqref{definition_of_derivative_of_Renyi_entropy} by means of finite differences on the lattice; focusing on the case in which the subsystem $A$ is a slab that is maximally extended in $D-2$ spatial directions and that has length $l$ in the direction orthogonal to the separation between $A$ and $B$, the entangling surface $\partial A$ does not depend on $l$ and the entropic c-function appearing in eq.~(\ref{definition_entropic_c-function}) is finite. Once the subsystem $A$ has been selected, say at Euclidean time $\tau$, all links emanating from sites belonging to $A$ in the Euclidean-time direction, connecting sites at time $\tau$ to sites at time $\tau+a$, are eliminated by setting the coupling $\beta^{(k,k)}=0$: the cut is now a $D-2$ dimensional slab, orthogonal to the time direction. Then links connecting the lower side of the cut in one replica to the upper side of the cut of the following one are introduced, and we set $\beta^{(k,k+1)}=\beta$. Using eq.~(\ref{definition_entropic_c-function}), the R\'enyi entropic c-function can then be written as
\begin{align}
\label{entropic_c-function_in_code}
C_n(l) = \frac{l^{D-1}}{|\partial A|}\frac{1}{n-1}\lim_{\epsilon\to 0}\frac{1}{\epsilon}\log\frac{Z_n(l)}{Z_n(l+\epsilon)}.
\end{align}
In practice, we set $\epsilon$ to the lattice spacing $a$ (the shortest distance that can be defined on the lattice) and evaluate the ratio of partition functions on the right-hand side of eq.~(\ref{entropic_c-function_in_code}) using eq.~(\ref{Jarzynski_theorem}), by ``shifting'' one of the two connected components of the entangling surface by $a$ during each non-equilibrium Monte~Carlo evolution of the system. This is done by linearly reducing the coupling of spins near the entangling surface with spins in another replica, while at the same time increasing their coupling to spins within the same replica, as sketched in fig.~\ref{fig:higher_dimensions_c-function_algorithm}. Note that in a $D$-dimensional lattice of spatial length $L$ the total number of links that are varied is $2L^{D-2}n$, regardless of the length of the cut.
%
%
%

Once one has calculated the derivative of the entanglement entropy for all the values of $l$, the R\'enyi entropy can be reconstructed (up to a non-universal constant) simply by numerical integration. This approach is known in literature as the increment trick and is used to improve results of simulations using reweighting~\cite{Alba:2011fu} and non-equilibrium techniques~\cite{Zhao:2021njg}. However our main interest in this work is to exploit this method, adapted to out-of-equilibrium simulations, to directly calculate the derivative of the R\'enyi entropies that, as we will discuss in section~\ref{sec:Ising_3D}, is a valuable tool to unveil universal, subleading corrections beyond the area law.

Notice also that the simplest way to perform the non-equilibrium evolution changing the length of $A$ is to vary all couplings in the same way: at time $t_m = \ti + m\frac{\tf-\ti}{N}$ all couplings at the edge of the cut take the value $\beta^{(k,k)}=\beta\left(1-\frac{m}{N}\right)$ and $\beta^{(k,k+1)}=\beta\frac{m}{N}$, but other protocols can be used as well. In section~\ref{app:a_different_protocol_for_the_three-dimensional_case} of the appendix, we discuss a different protocol, which is inspired by recent lattice calculations of the entanglement entropy in gauge theories~\cite{Rindlisbacher:2022bhe, Jokela:2022fvh, Jokela:2023yun}.

In order to enhance the computational speed of our algorithm, we exploited the higher level of parallelization achievable by using graphics processing units. To this purpose we adapted an existing CUDA code~\cite{Komura:2012gbs, Komura:2014cpf}, which implements a parallelized version of Swendsen-Wang cluster updates~\cite{Swendsen:1987ce} using two different cluster-labeling algorithms~\cite{Hawick:2010pgc, Kalentev:2010ccl}, to efficiently simulate the $2D$ and $3D$ Ising models. We suitably modified the code to implement the replica geometry and Jarzynski's algorithm. The parallelization achievable by exploiting GPUs is significant, since we were able to parallelize not only the updating of all the spins on a replica system, but also the non-equilibrium trajectories.

\noindent
%
%

\section{Results for the Ising model in two dimensions}
\label{sec:Ising_2D}

In the present section we present the results of our simulations in a very simple model, for which the analytical form of the entropic c-function is known, namely the Ising model in two dimensions.

In $D=2$ the model defined in eq.~(\ref{Ising_model}) was first solved exactly by Onsager~\cite{Onsager:1943jn} and has since been solved again, using different techniques, by various other authors~\cite{Kaufman:1949ks, Kac:1952tp, Hurst:1960oae, Schultz:1964fv}, becoming a sort of ``archetype'' for integrable models~\cite{Yang:1967bm, Baxter:1972hz}.

At the critical point of the theory, corresponding to $\beta_c = \frac{1}{2}\log\left(1+\sqrt{2}\right)$, the entropic c-function is expected to be described by eq.~(\ref{entropic_c-function_on_the_cylinder_2D}) with $c=\frac{1}{2}$, up to finite size corrections that we will discuss in what follows. Moreover, in this work we focus on the calculation of the entropic c-function associated to the second R\'enyi entropy, $C_2$, since it requires a  lower computational effort compared to higher values of $n$.

The conventions and notations we use in this section are as follows.
\begin{itemize}
\item $L_\tau$ denotes the extent of the lattice in the Euclidean-time direction, while $L$ denotes the size of the lattice in the spatial direction. 
Since we are interested in the behavior of the entanglement in a quantum system at zero temperature, we work with lattices where $L_\tau \gg L$, in particular $L_{\tau} = 8 L$ (we will discuss this choice in what follows).
\item $l$ denotes the length of the cut; we also define a normalized length as $x = \frac{l}{L}$.
\item Periodic boundary conditions are imposed in all directions.
\item $N$ is the number of discrete time intervals in which the out-of-equilibrium evolution is divided, while $n_t$ is the number of trajectories used to calculate the average in eq.~(\ref{Jarzynski_theorem}).
\item All data have been extracted after an initial thermalization of $\Ntherm = 1000$ Monte Carlo steps; we estimated the integrated autocorrelation time at the critical point to be of the order of a few of our update steps.
\end{itemize}

\subsubsection*{Direct and reverse protocols}
%
%
\begin{figure}[t]
\centering
\begin{subfigure}{.45\textwidth}
\includegraphics[width=\textwidth]{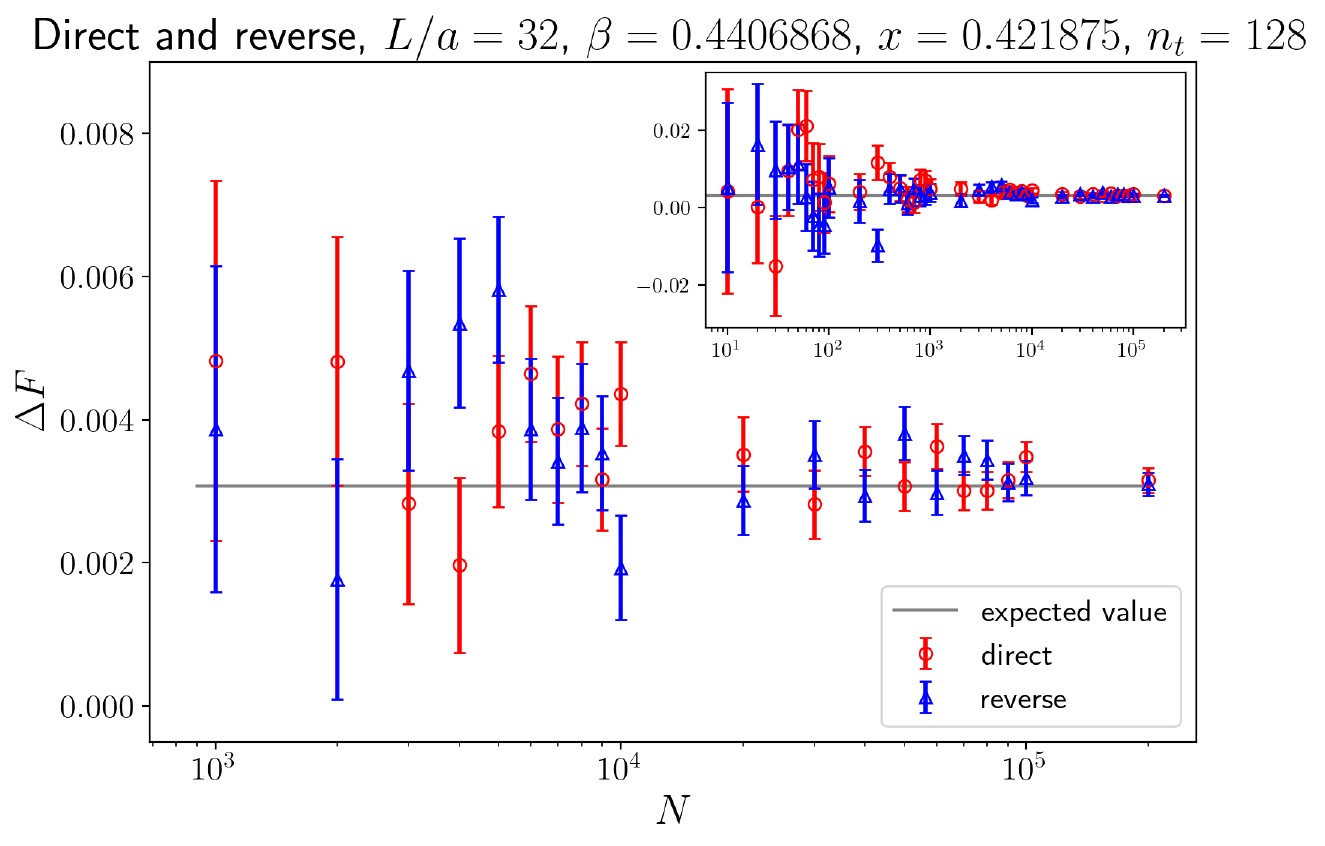}
\caption{}
\label{subfig:direct_reverse_2d_a}
\end{subfigure}
\begin{subfigure}{.45\textwidth}
\includegraphics[width=\textwidth]{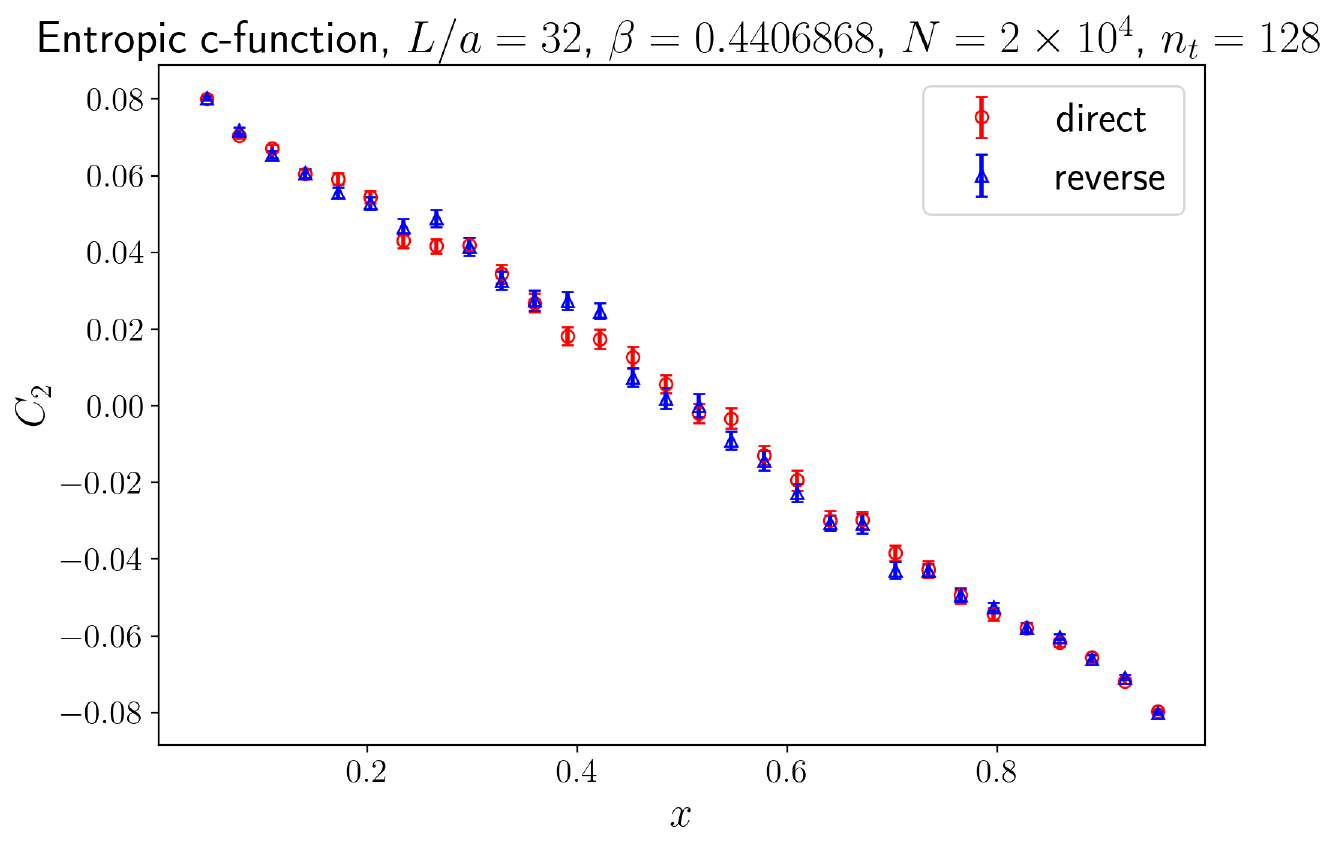}
\caption{}
\label{subfig:direct_reverse_2d_b}
\end{subfigure}
\begin{subfigure}{.45\textwidth}
\includegraphics[width=\textwidth]{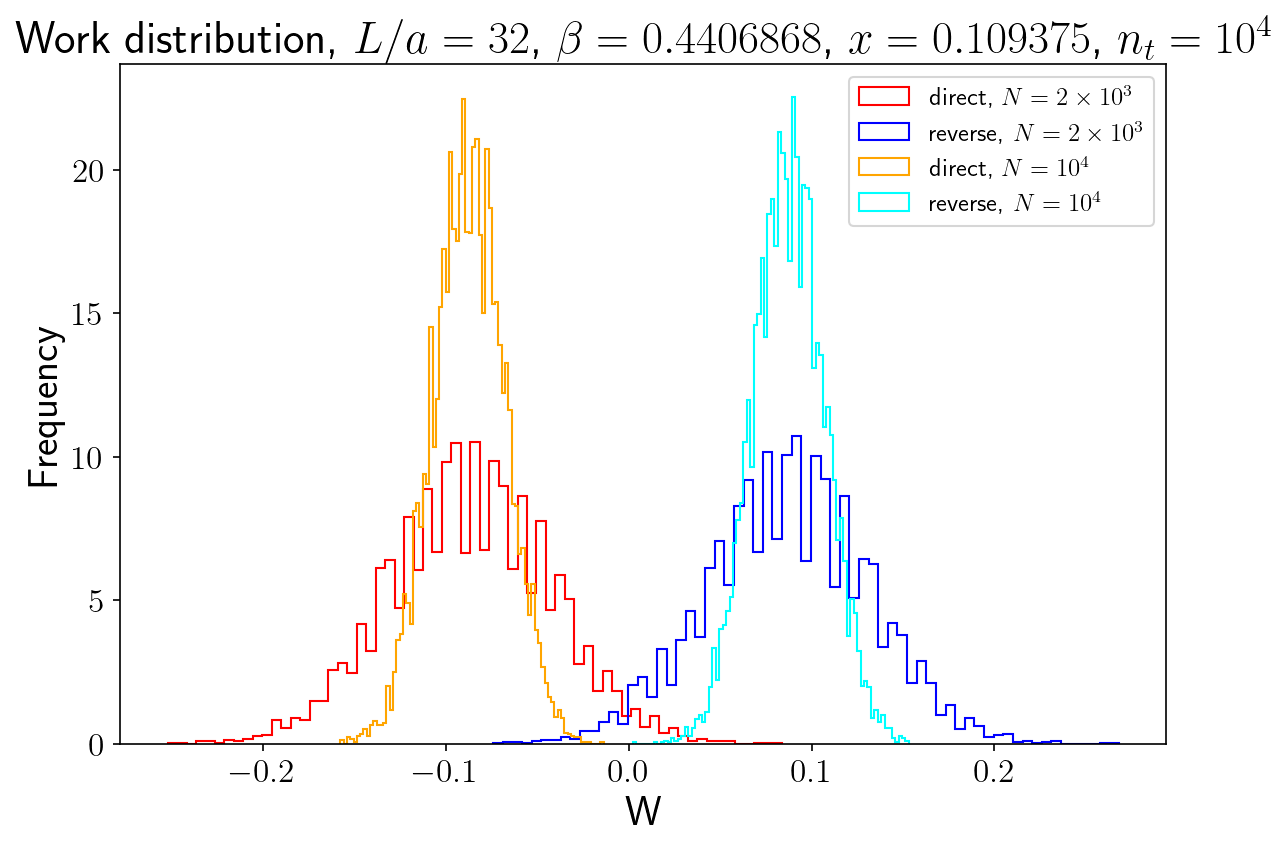}
\caption{}
\label{subfig:direct_reverse_2d_c}
\end{subfigure}
\begin{subfigure}{.45\textwidth}
\includegraphics[width=\textwidth]{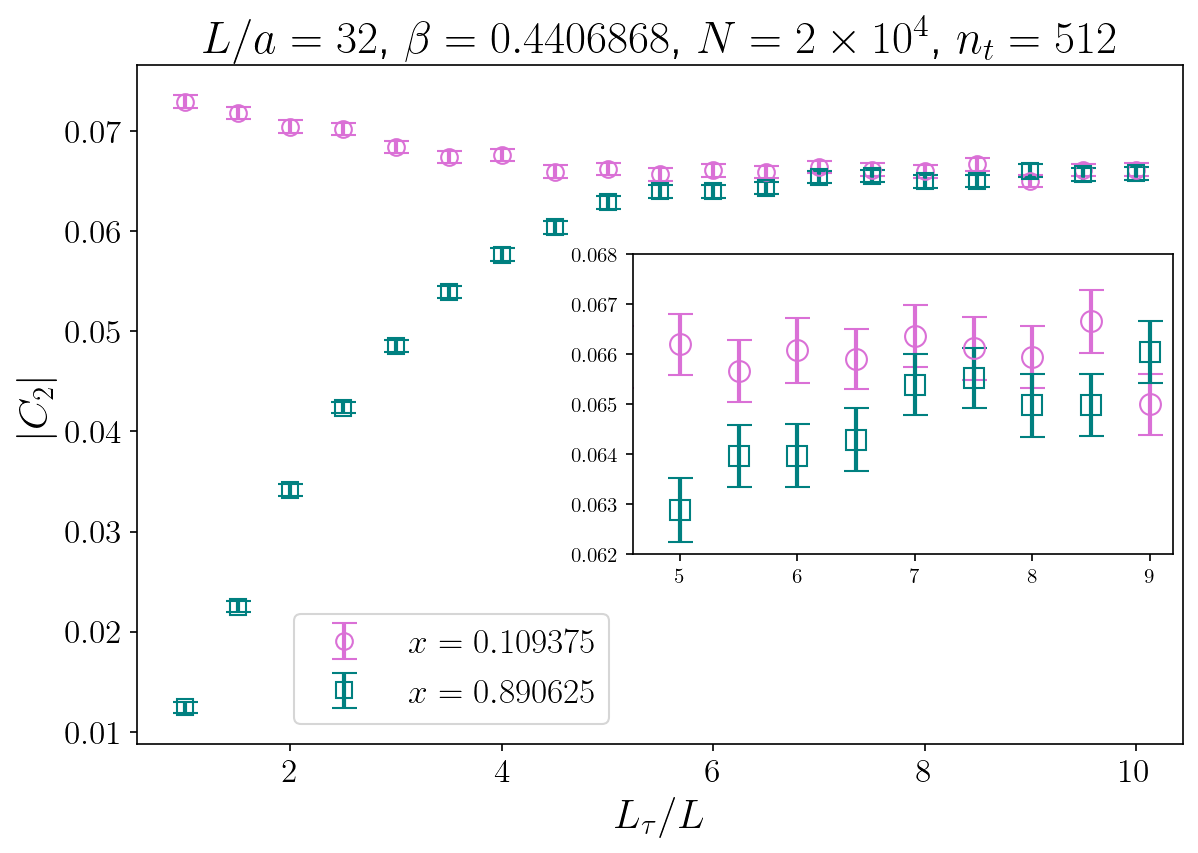}
\caption{}
\label{subfig:temperature_2d_d}
\end{subfigure}
\caption{Consistency checks of Jarzynski's algorithm: (a) direct and reverse protocols at different number of Monte Carlo steps (with the inset showing a wider range of $N$); (b) direct and reverse calculation of the entropic c-function; (c) distribution of the total off-equilibrium work from eq.~(\ref{off_equilibrium_work}). Check of the zero-temperature condition: (d) the quantum system is at zero temperature when the entropic c-function changes its sign under the exchange $x\rightarrow 1-x$; inset: zoom on the region where the two data sets converge.}
\end{figure}\noindent
%
%
A reliable way to test the consistency of Jarzynski's algorithm is to compare direct and reverse protocols. From now on, the ``direct'' realization will be the one in which we calculate the backward derivative (i.e., we start with a cut of length $l$ and end with a cut of length $l-a$), while the ``reverse'' realization will be the one defined in terms of the forward derivative (i.e., from $l-a$ to $l$).

Our data suggest that the average appearing in eq.~(\ref{Jarzynski_theorem}) is never biased by poor statistics; this is not a surprise, since in two dimensions and in a two-replica geometry the total number of couplings that are varied is equal to $4$, regardless of the size of the lattice, hence we expect the system to be driven slowly out of equilibrium for almost all values of $N$. This is confirmed by the results shown in fig.~\ref{subfig:direct_reverse_2d_a}, where direct and reverse protocols are compared for a range of values of $N$ at fixed $n_t$. After a first range of values, between $10$ and $100$, in which results are dominated by statistical noise, direct and reverse protocols become consistent with each other for almost all values of $N$, up to some slight discrepancies that are compatible with statistical fluctuations.

In fig.~\ref{subfig:direct_reverse_2d_b} a further comparison of the direct and reverse implementations of the algorithm is shown, between two set of data at fixed $N$ and $n_t$ and with different values of $x$. The reduced $\chi^2$ is found to be $\frac{\chi^2}{\nu}=1.37$, showing a good match between direct and reverse realizations.

Finally, fig.~\ref{subfig:direct_reverse_2d_c} shows the work distributions for different values of $N$ and for direct and reverse protocols, which are approximately symmetric, as expected from previous work~\cite{Francesconi:2020fgi}. Note that, as $N$ increases, the distributions shrink, since one is getting closer to a quasi-static evolution and, as a consequence, statistical fluctuations are reduced; therefore, differently from a common Monte Carlo simulation, in this case we have two parameters to control the statistical uncertainty of our data, namely $N$ and $n_t$. Since the algorithm seems to be consistent for a large range of $N$, how to tune $N$ and $n_t$, given the desired precision, is dictated by numerical convenience, depending also on the available computing resources. In our implementation, since all trajectories are parallelized, increasing $n_t$ means increasing the memory allocation on the GPU, while increasing $N$ translates into a longer computing time.

\subsubsection*{Zero-temperature condition}

In this work we are interested in studying the entanglement for systems at zero temperature, a condition that, on the lattice, is satisfied if $L_{\tau}$ is sufficiently large. The observable itself we are considering provides a way to quantitatively check if the condition is satisfied. We already mentioned that the R\'enyi entropies are symmetric at zero temperature, eq.~(\ref{symmetry_of_Renyi_at_zero_temperature}). This property is inherited by the derivative, and, hence, by the symmetrized entropic c-function, which satisfies
\begin{align}
C_n(x) = -C_n(1-x).
\end{align}
Fig.~\ref{subfig:temperature_2d_d} shows two complementary values of the entropic c-function at different values of the $\frac{L_{\tau}}{L}$ ratio. Error bars take into account both statistical and systematic uncertainties. We find that for $\frac{L_{\tau}}{L}\gtrsim 7$ the data are compatible within their uncertainties.

\subsubsection*{Entropic c-function at the critical point}
%
%
\begin{figure}[t]
\centering
\begin{subfigure}{.45\textwidth}
\includegraphics[width=\textwidth]{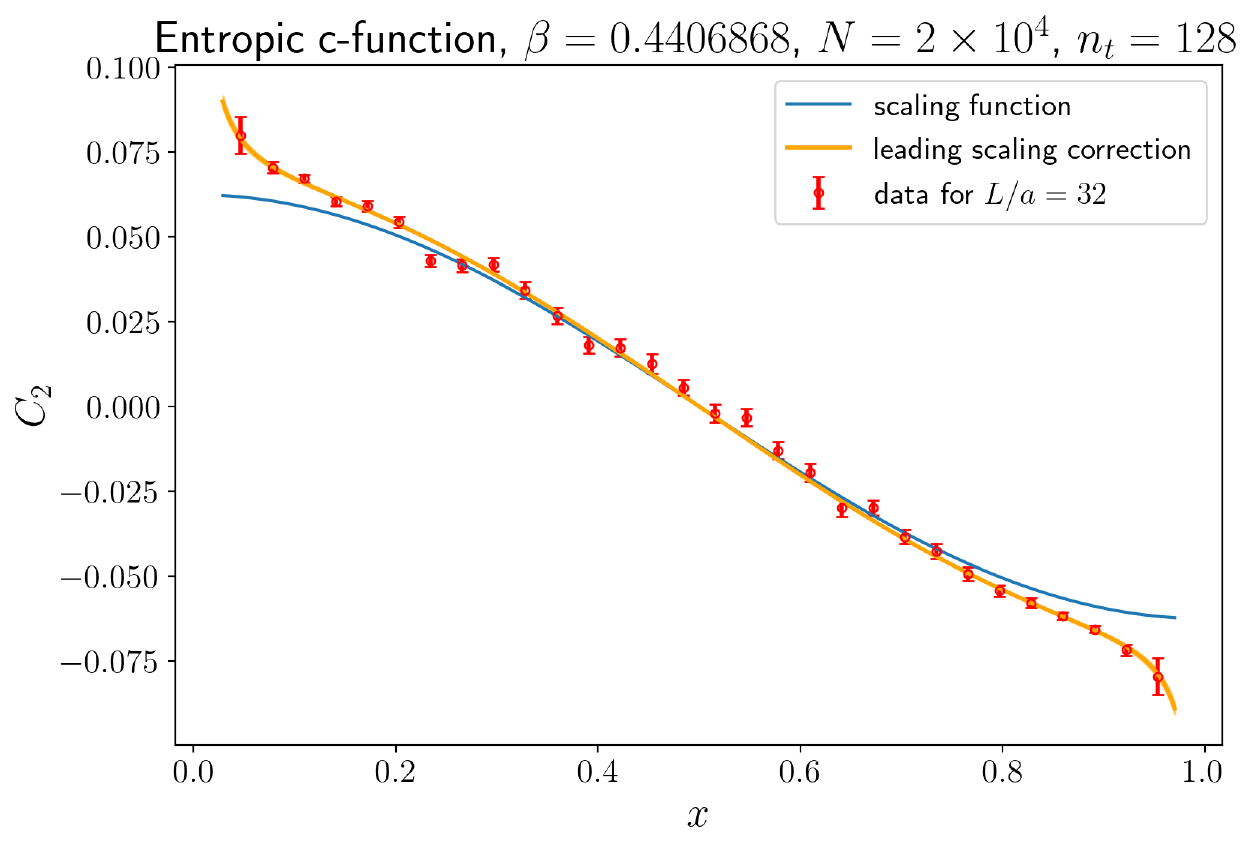}
\caption{}
\label{subfig:fit_cfun_32_2d}
\end{subfigure}
\begin{subfigure}{.45\textwidth}
\includegraphics[width=\textwidth]{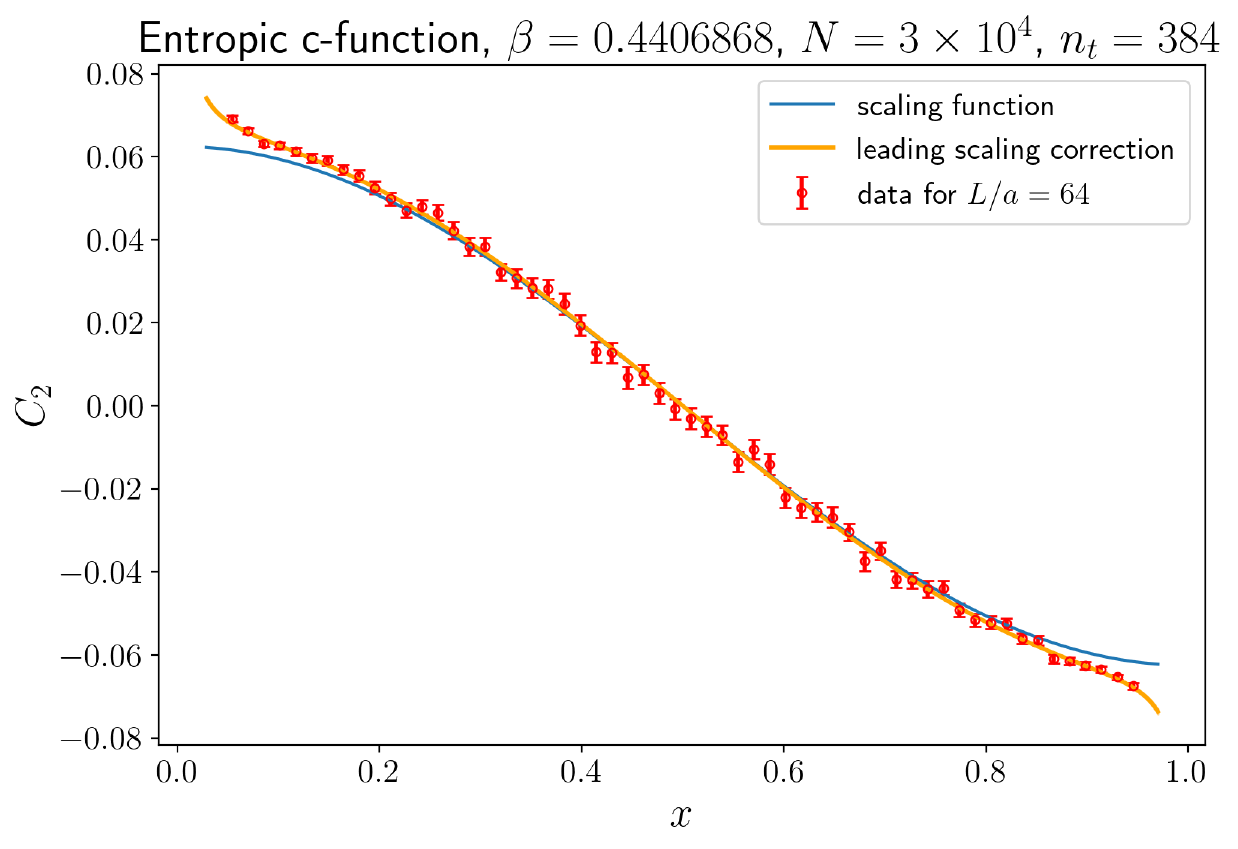}
\caption{}
\label{subfig:fit_c-fun_64_2d}
\end{subfigure}
\begin{subfigure}{.45\textwidth}
\includegraphics[width=\textwidth]{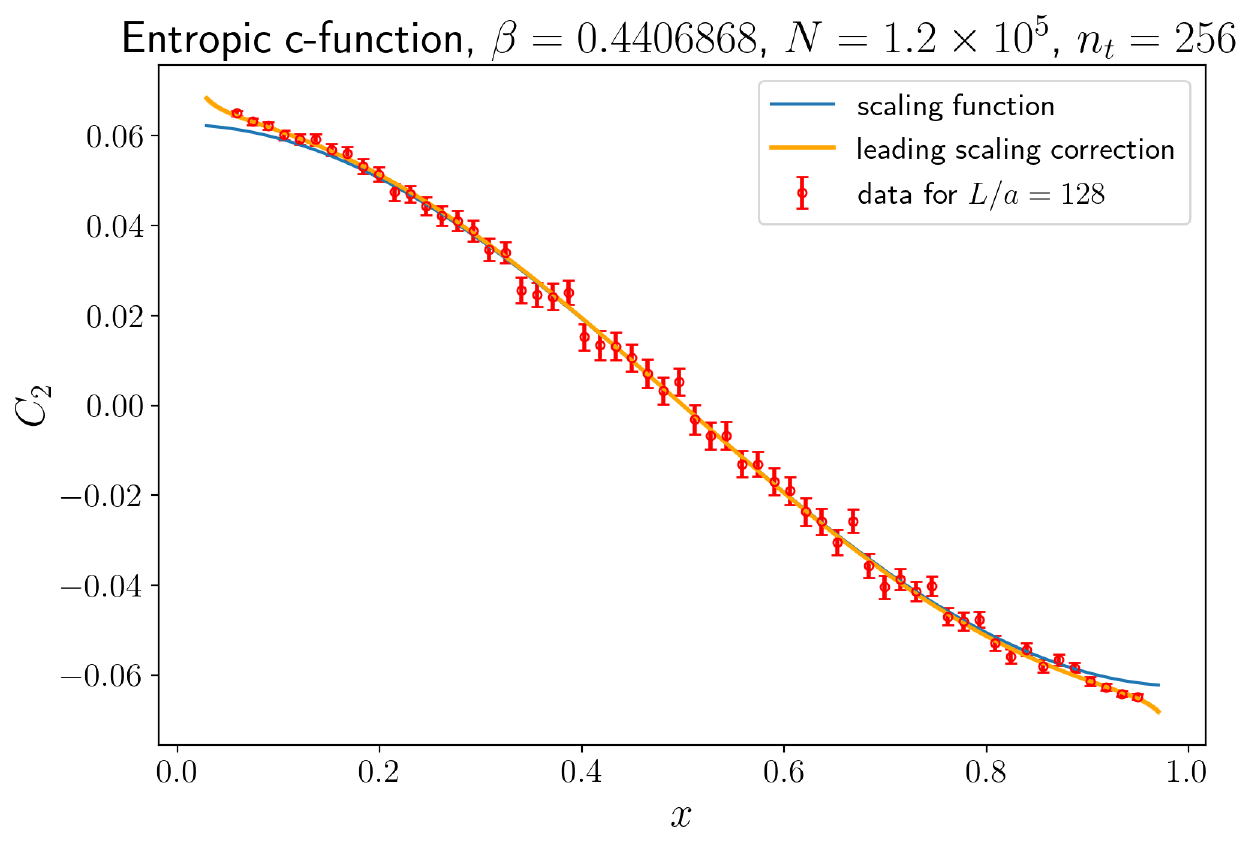}
\caption{}
\label{subfig:fit_c-fun_128_2d}
\end{subfigure}
\begin{subfigure}{.45\textwidth}
\includegraphics[width=\textwidth]{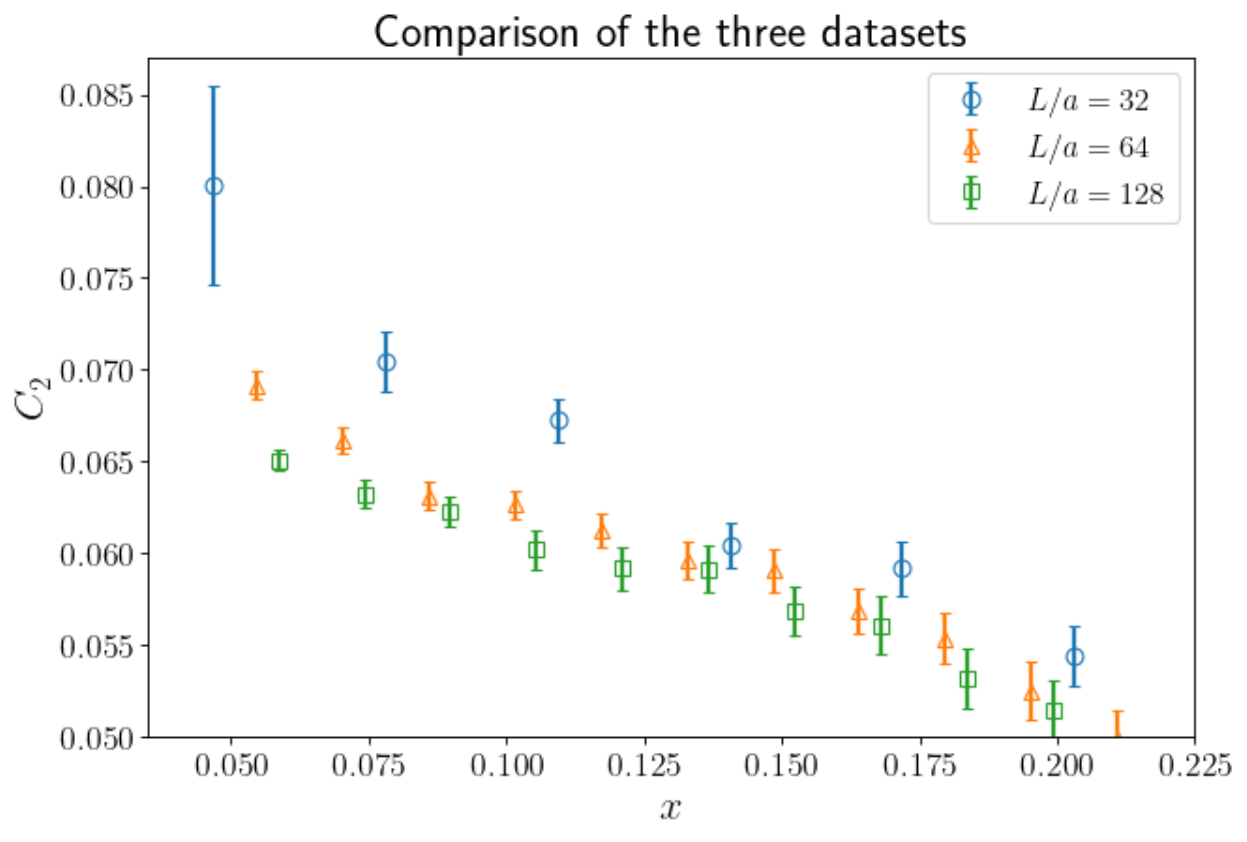}
\caption{}
\label{subfig:comparison_c-fun_2d}
\end{subfigure}
\caption{First three plots: entropic c-function at different values of $L$ compared with the CFT result in the scaling limit and the first scaling correction. Last plot: comparison of the three datasets for small values of $x$.}
\label{fig:fit_c-fun_2d}
\end{figure}\noindent
%
%
Our numerical results for the $D=2$ Ising model can be directly compared with the CFT result, eq.~(\ref{Renyi_entropy_on_a_cylinder}). However this result is valid only when $L$ and $l$ are much larger than the lattice spacing $a$, otherwise finite-size corrections have to be taken into account. The theory of corrections to scaling for the R\'enyi entropies in $1+1$ dimensions has been developed in ref.~\cite{Cardy:2010zs}, where it was shown that the leading correction scales as $\order{l^{-2h/n}}$, where $n$ is the number of replicas and $h$ the conformal dimension of a relevant operator localized at the endpoints of the cut. In ref.~\cite{Xavier:2011np} it was shown numerically that, for a large class of theories, $h$ is the conformal dimension of the energy operator. For the Ising model in an $n=2$ replica geometry we then expect a correction of order $\order{l^{-1}}$. More precisely, the fit function that we used to analyze our data is
\begin{align}
f(x;k) = \frac{1}{16}\cos(\pi x) + \frac{k}{2 L}\cot(\pi x),
\label{fit_function_2d}
\end{align}
where the first term on the right-hand side is the scaling function and the second one is the scaling correction, multiplied by a non-universal constant $k$. Figure~\ref{fig:fit_c-fun_2d} shows data for $\frac{L}{a}=32$, $64$ and $128$, the scaling function (blue curve) and the fit function (orange curve). The results are summarized in table~\ref{tab:data_scaling-function_comparison_2d}.
\begin{table}[H]
\begin{center}
\begin{tabular}{|c|c|c|}
\hline
$\frac{L}{a}$ & $\frac{\chi^2}{\nu}$ & $\frac{\chi^2}{\nu-4}$ \\
\hline \hline
$32$ & $9.65$ & $7.17$ \\
\hline
$64$ & $7.17$ & $2.99$ \\
\hline
$128$ & $2.83$ & $1.08$ \\
\hline
\end{tabular}
\begin{tabular}{|c|c|c|}
\hline
$\frac{L}{a}$ & $\frac{\chi^2}{\nu}$ & $k$ \\
\hline \hline
$32$ & $1.30$ & $0.162(10)$ \\
\hline
$64$ & $0.77$ & $0.138(7)$ \\
\hline
$128$ & $0.69$ & $0.143(13)$ \\
\hline
\end{tabular}
\end{center}
\caption{Left-hand side: comparison between our data and the scaling function. Right-hand side: comparison between our data and the fit function~\eqref{fit_function_2d}.}\label{tab:data_scaling-function_comparison_2d}
\end{table}\noindent
Firstly we compared our data with the scaling function only, i.e., the first term of eq.~\eqref{fit_function_2d}, and we collected the $\chi^2$ values in the table on the left. The mid column refers to the whole sets of data displayed in fig.~\ref{fig:fit_c-fun_2d}, while on the right we excluded the first two and the last two points of each set, as they are those that are most affected by finite-size effects. As expected, data converge to the scaling curve as $L$ grows, and in particular we obtain a perfect match for $\frac{L}{a}=128$ if we exclude the data at the boundaries of the interval under consideration.

In the table on the right, the $\chi^2$ and the best-fit results are displayed, showing perfect agreement between the theoretical predictions and our data. 

\subsubsection*{Variation with $\beta$}
%
%
\begin{figure}[t]
\centering
\begin{subfigure}{.48\textwidth}
\includegraphics[width=\textwidth]{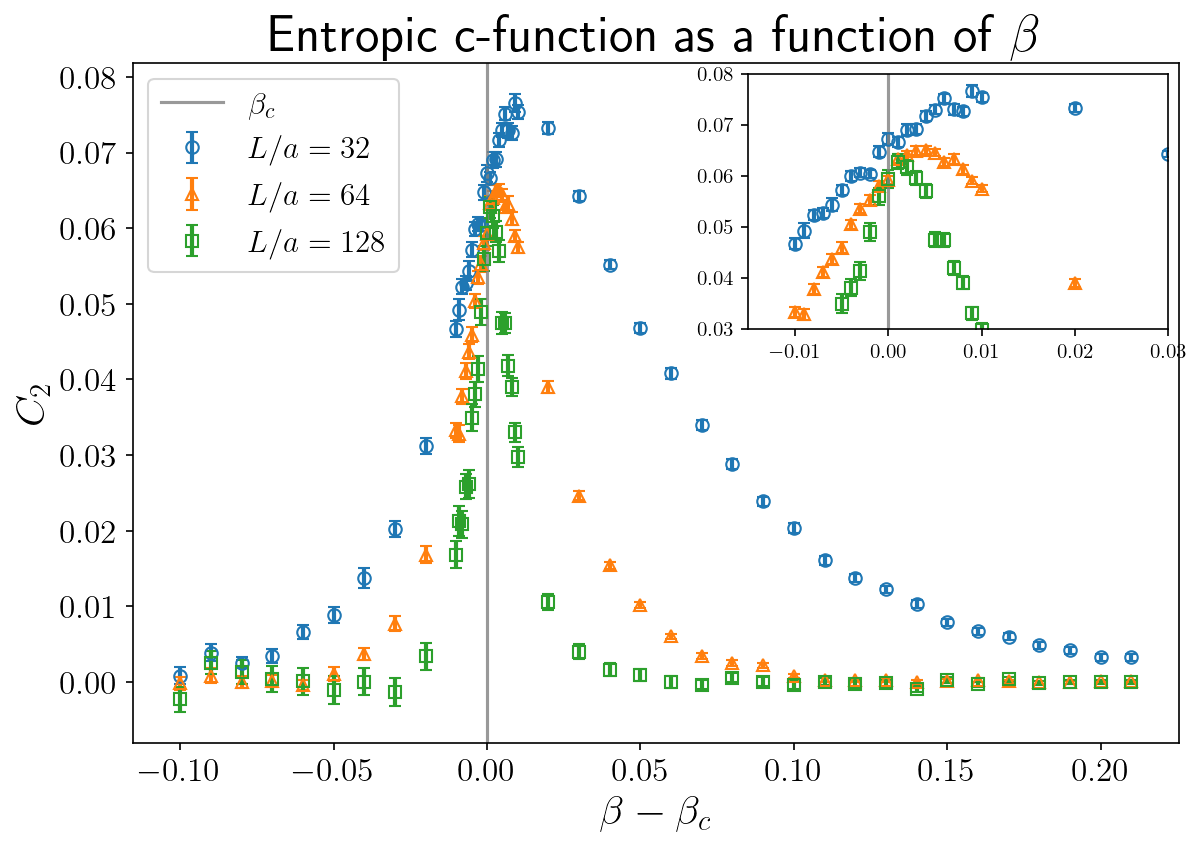}
\caption{}
\label{subfig:beta_2d}
\end{subfigure}
\begin{subfigure}{.48\textwidth}
\includegraphics[width=\textwidth]{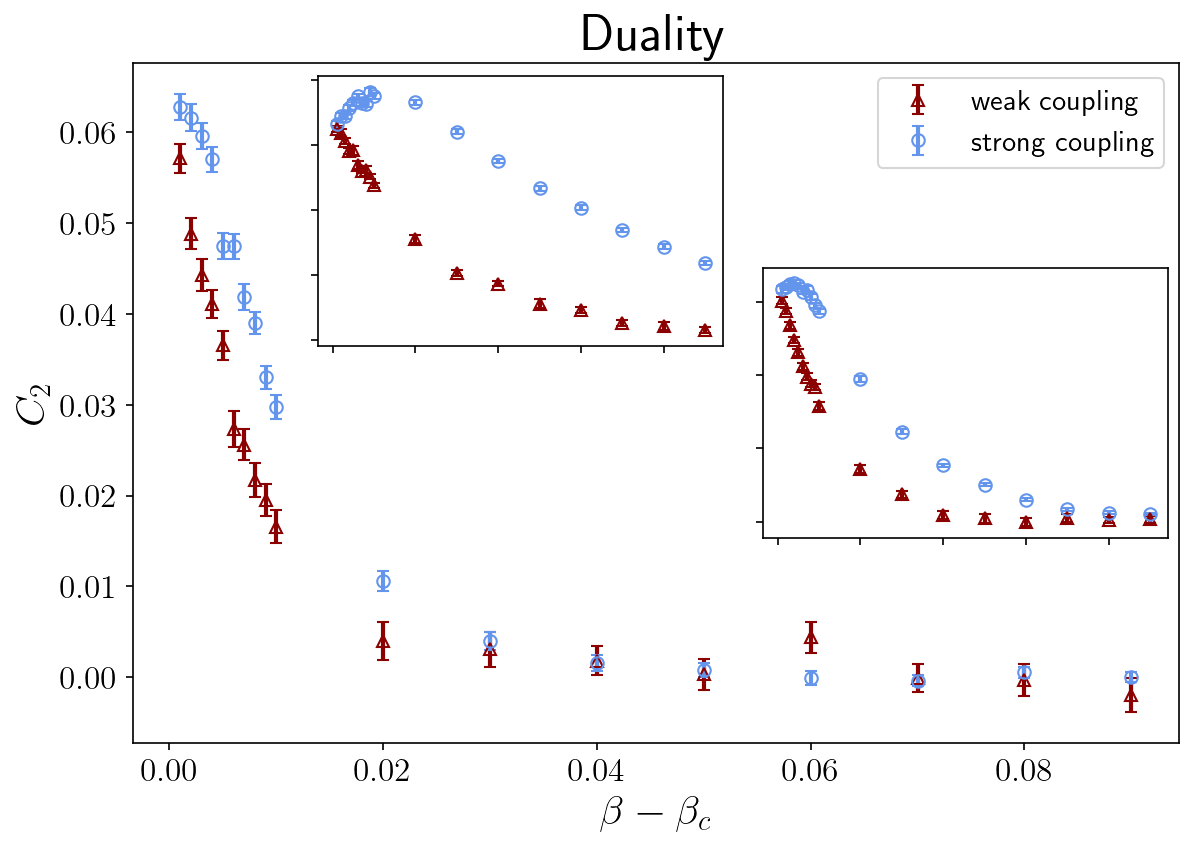}
\caption{}
\label{subfig:duality_2d}
\end{subfigure}
\caption{(a): $C_2$ as a function of $\beta$ for $\frac{L}{a}=32$, $64$, and $128$ and for $\frac{l}{a}=4$, $8$, and $16$ respectively; $N$ and $n_t$ are the same as in fig.~\ref{fig:fit_c-fun_2d}; the inset shows a zoom on the critical region. (b): comparison between data at $\beta > \beta_c$ and their dual at $\beta < \beta_c$ for $\frac{L}{a}=32$ (left inset), $\frac{L}{a}=64$ (right inset) and $\frac{L}{a}=128$ (main plot).}
\label{fig:beta_2d}
\end{figure}\noindent
%
%
We conclude this section with a study of the behavior of the entropic c-function as the coupling $\beta$ is varied. This quantity is expected to have a maximum at the critical point and to be monotonically decreasing when one moves away from criticality~\cite{Casini:2004bw, Casini:2006es}. Our data, shown in fig.~\ref{subfig:beta_2d}, are in agreement with this prediction up to finite-size effects, that are pronounced for $\frac{L}{a}=32$ and decrease as $L$ grows.

We also studied the behavior of the entropic c-function under the self-dual map of the Ising model~\cite{Radicevic:2016tlt}. It is well known that in two dimensions the partition function of the Ising model at $\beta > \beta_c$ can be related to the one of the same model at $\beta^\star < \beta_c$ via the relation~\cite{Kramers:1941kn}
\begin{align}
\exp\left(-2\beta\right)=\tanh\beta^\star.
\end{align}
It is important to note that the self-duality strictly holds only in the infinite-volume limit. Indeed from fig.~\ref{subfig:duality_2d} it is evident that the two data sets have different behavior at small $L$, while they tend to converge for larger $L$. However the convergence is slow and even for $\frac{L}{a}=128$ the reduced $\chi^2$ is large, $\frac{\chi^2}{\nu}\simeq24$. It is also important to realize that in a replica geometry self-duality is also violated by the presence of the cut. As described in ref.~\cite{Cardy:2010zs}, since the cut lies along the links of the dual lattice, the dual spins located at the edges of the cut have a larger number of nearest neighbors: for a square lattice the nearest neighbors are $4n$, which induces an explicit local breakdown of the self-duality of the model. However, it is beyond the goals of this work to analyze the effects of the presence of the cut on the duality properties of the Ising model, and we leave the study of this issue for future work.

\section{Results for the Ising model in three dimensions}
\label{sec:Ising_3D}

%
%
\begin{figure}[t]
\centering
\begin{subfigure}{.45\textwidth}
\includegraphics[width=\textwidth]{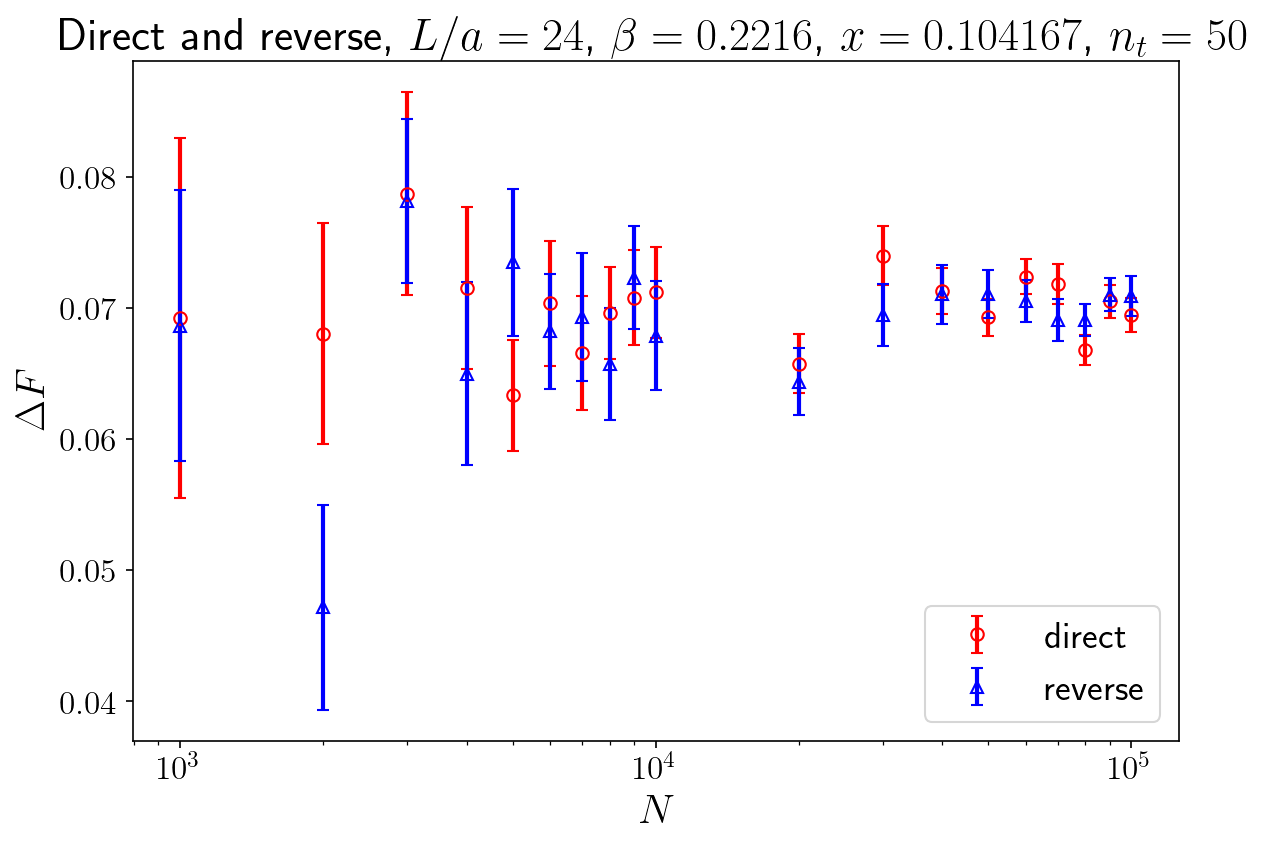}
\caption{}
\label{subfig:direct_reverse_3d_a}
\end{subfigure}
\begin{subfigure}{.45\textwidth}
\includegraphics[width=\textwidth]{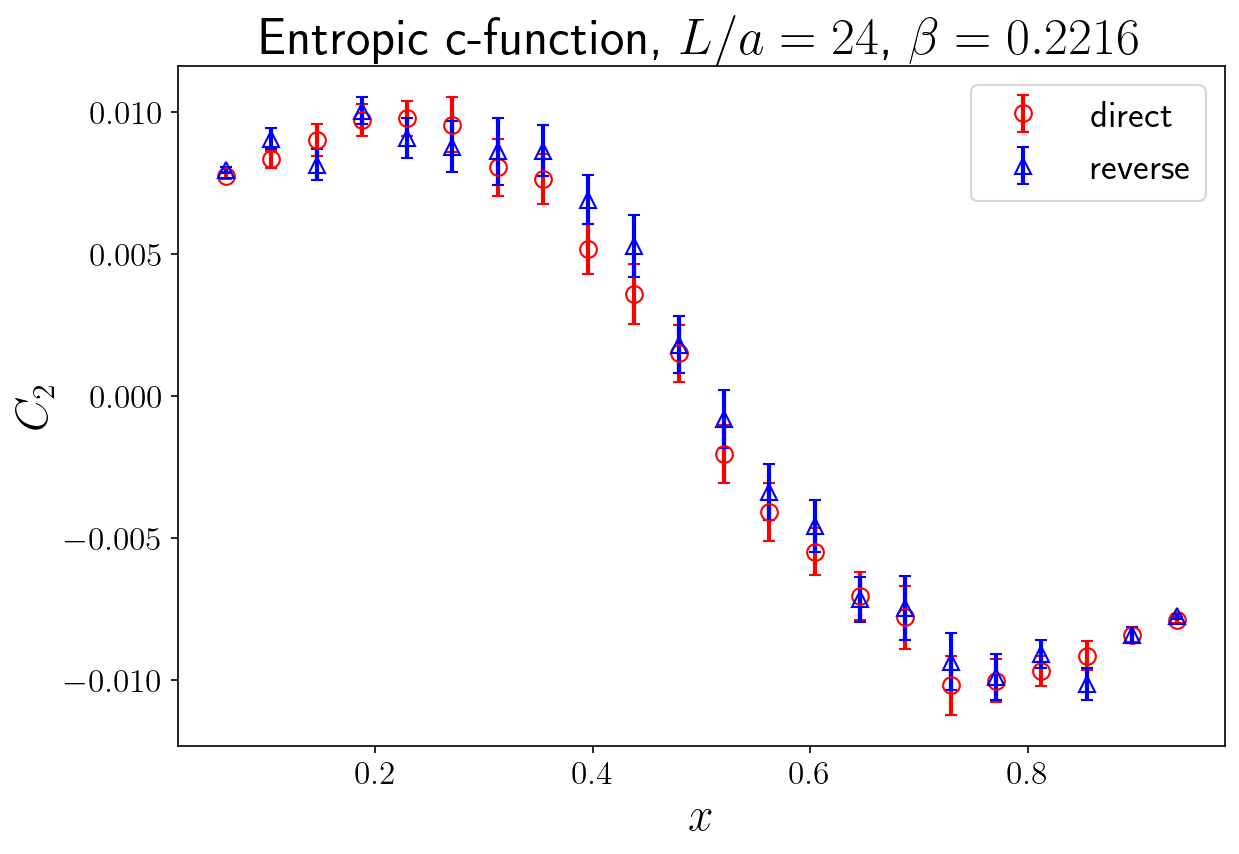}
\caption{}
\label{subfig:direct_reverse_3d_b}
\end{subfigure}
\begin{subfigure}{.45\textwidth}
\includegraphics[width=\textwidth]{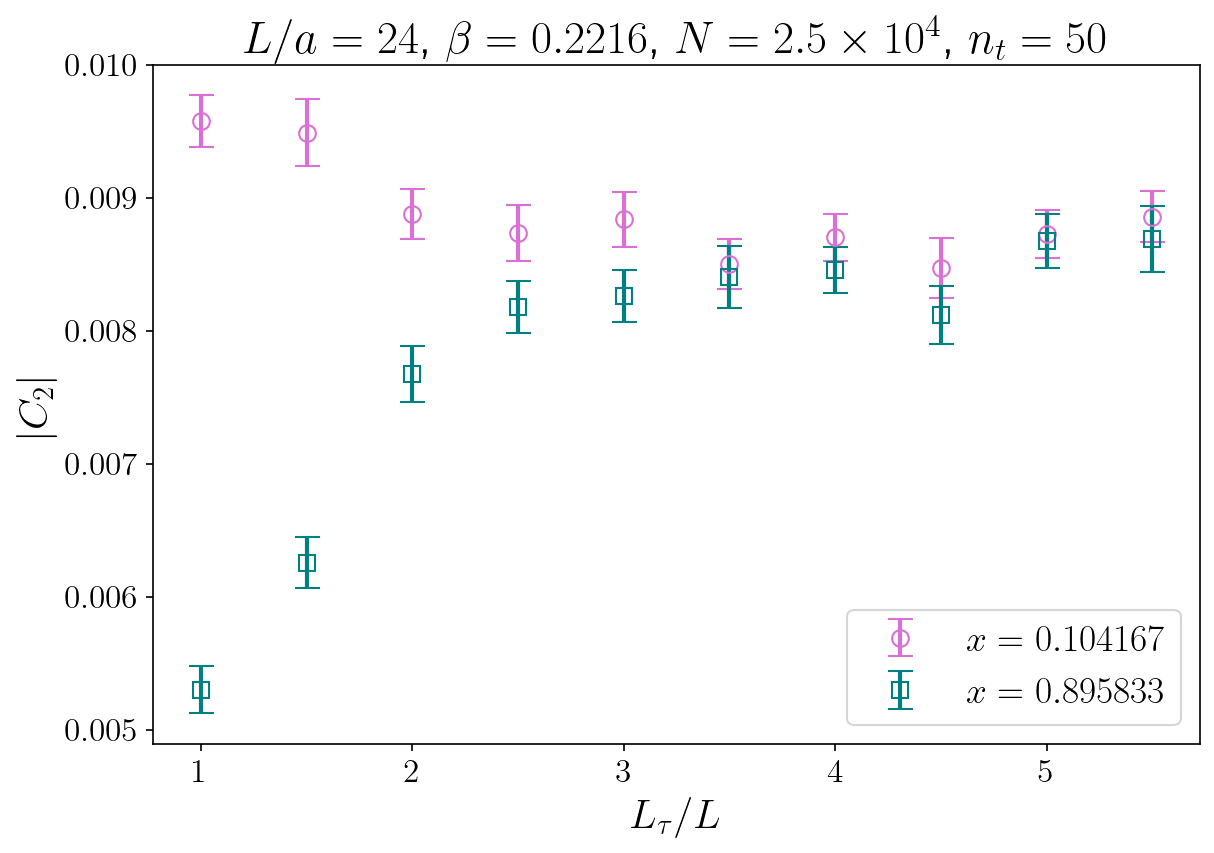}
\caption{}
\label{subfig:temperature_3d_c}
\end{subfigure}
\caption{Consistency checks of Jarzynski's algorithm: (a) direct and reverse protocols for a different number of Monte Carlo steps. (b) Direct and reverse calculation of the entropic c-function; $N$ and $n_t$ span from $2.5\times 10^4$ and $50$ respectively for values of $x$ close to $0$ and $1$, to $10^5$ and $100$ for values of $x$ close to $\frac{1}{2}$. (c) Zero-temperature condition.}
\end{figure}\noindent
%
%
In this section we present our results for the Ising model in three dimensions. In contrast with its two-dimensional counterpart, the three-dimensional Ising model still defies all attempts towards a complete solution -- and, in particular, no analytical result on its entanglement content is known. In spite of (or, rather, by virtue of) its simplicity, the model remains relevant for a wide variety of problems, including in high-energy physics (see, e.g., ref.~\cite[section~5]{Aarts:2023vsf}). While it is generally believed that in the continuum limit this model should admit a superstring description~\cite{Fradkin:1980gt, Polyakov:1981re, Casher:1981hc, Sedrakian:1982ne, Itzykson:1982ec, Kavalov:1986xw, Dotsenko:1986mv, Dotsenko:1986ur, Polyakov:1987ez, Kavalov:1987vm, Ambjorn:1992nq, Iqbal:2020msy}, this intuition has not allowed one to derive explicit results (and, in fact, it has been pointed out that it entails some non-trivial aspects~\cite{Distler:1992rr}). At present, much numerical information about the three-dimensional Ising model is known, including, in particular, very precise estimates of the critical exponents~\cite{El-Showk:2012cjh, El-Showk:2014dwa, Gliozzi:2014jsa} based on the conformal-bootstrap approach~\cite{Rattazzi:2008pe}.

Also for the three-dimensional Ising model, that we studied using the same algorithm discussed in section~\ref{sec:Ising_2D}, after consistency checks of Jarzynski's algorithm, we focused our attention on the behavior of the entropic c-function at the critical point of the theory $\beta_c=0.221654626(5)$~\cite{Ferrenberg:2018zst} as a function of $x$, as well as the behavior when $\beta$ is varied.

The same conventions of the previous section are used here; moreover, we take the lattice extent to be the same in the two spatial directions, $L_x = L_y = L$, while the lattice extent in the remaining, Euclidean-time, direction is $L_{\tau} = 4L$.

\subsubsection*{Direct and reverse protocols}

Also in the three-dimensional case our results seem never to be biased by poor statistics. It should be noted that, although in this case the number of couplings which are varied is proportional to the spatial extent of the lattice, the relative number, compared to the size of the system, scales as $(L L_{\tau})^{-1}$, and this is true in all dimensions.

Figure~\ref{subfig:direct_reverse_3d_a} and fig.~\ref{subfig:direct_reverse_3d_b} show that direct and reverse protocols give the same result; in particular the reduced $\chi^2$ for the two sets of data in fig.~\ref{subfig:direct_reverse_3d_b} is $\frac{\chi^2}{\nu}=0.70$.

\subsubsection*{Zero-temperature condition}

In fig.~\ref{subfig:temperature_3d_c} two complementary values of the cut are shown. The two datasets start converging from $\frac{L_\tau}{L}\simeq 3.5$: this motivates our choice $L_\tau = 4L$.

\subsubsection*{Entropic c-function at critical point}
%
%
\begin{figure}[t]
\centering
\begin{subfigure}{.45\textwidth}
\includegraphics[width=\textwidth]{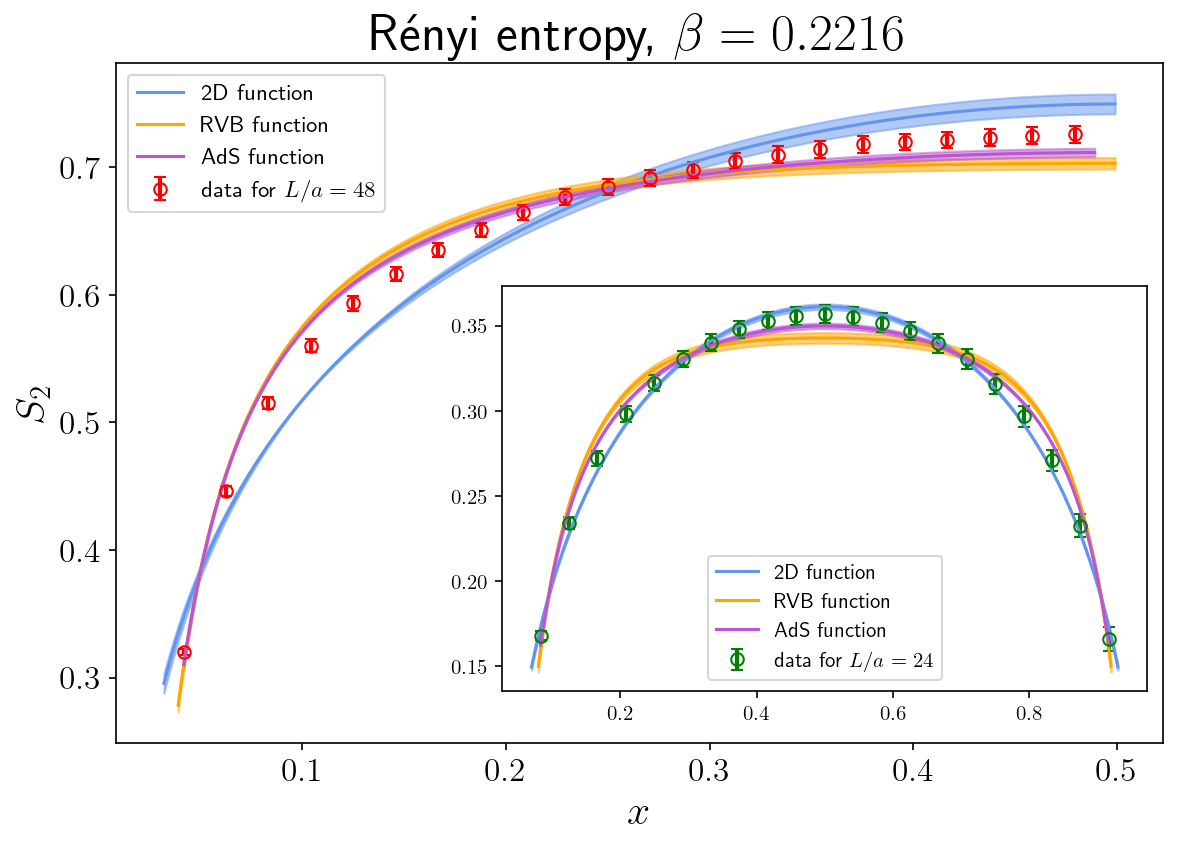}
\caption{}
\label{subfig:fit_entropy_3d}
\end{subfigure}
\begin{subfigure}{.45\textwidth}
\includegraphics[width=\textwidth]{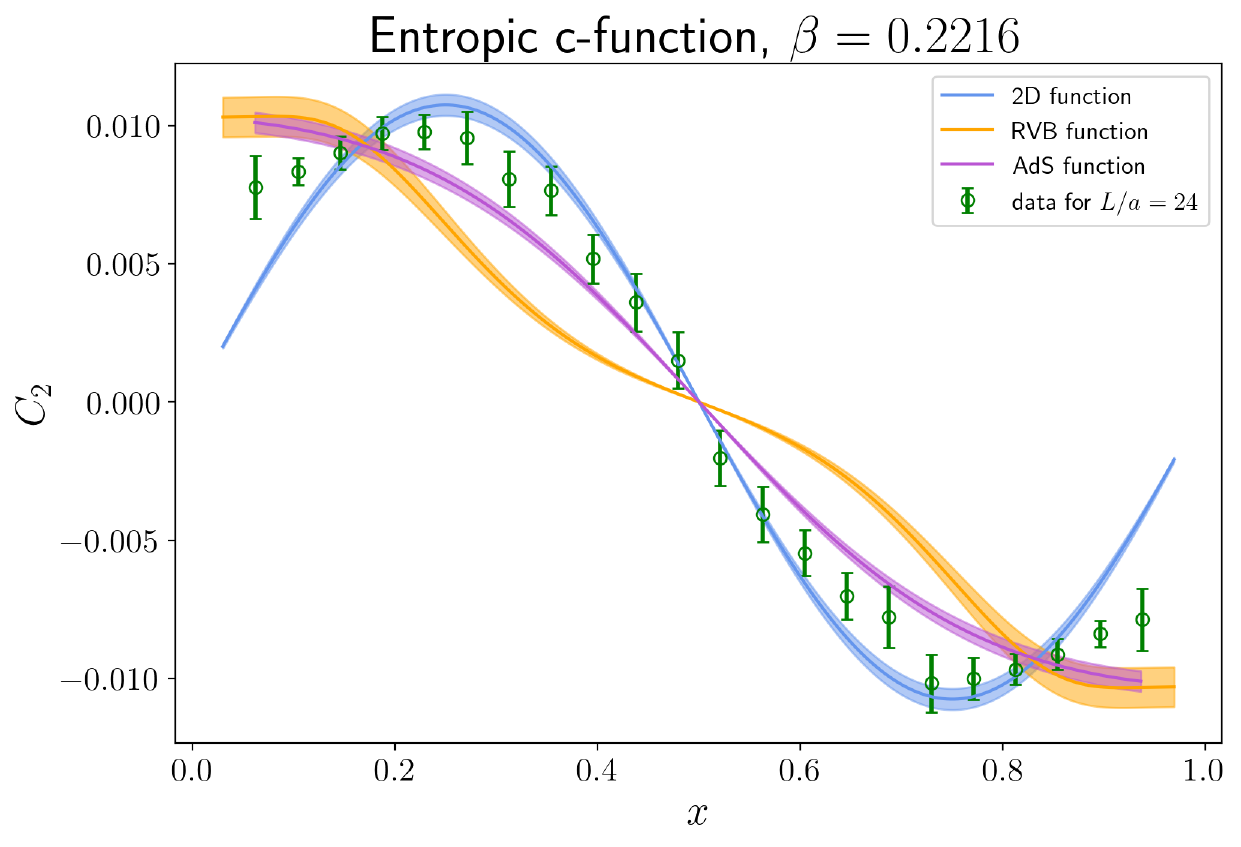}
\caption{}
\label{subfig:fit_c-fun_24_3d}
\end{subfigure}
\begin{subfigure}{.45\textwidth}
\includegraphics[width=\textwidth]{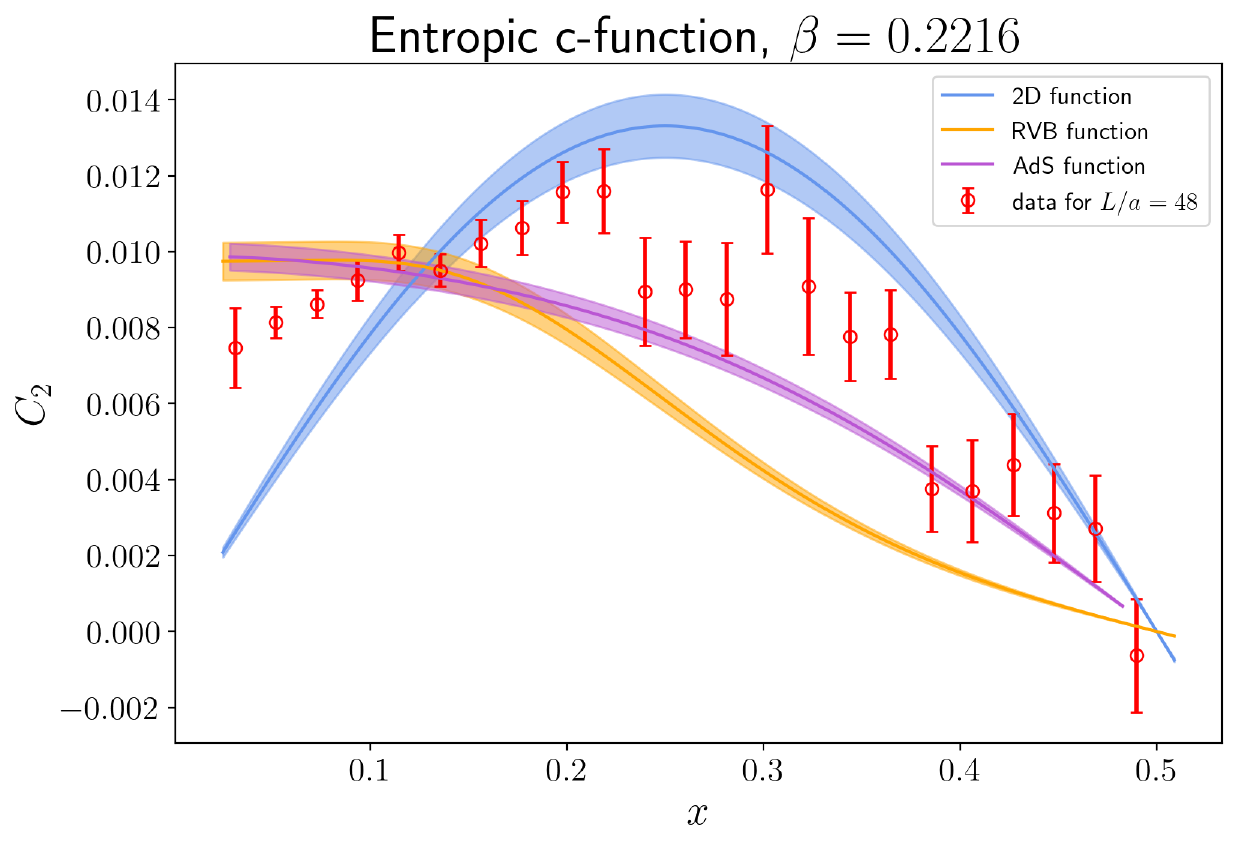}
\caption{}
\label{subfig:fit_c-fun_48_3d}
\end{subfigure}
\begin{subfigure}{.45\textwidth}
\includegraphics[width=\textwidth]{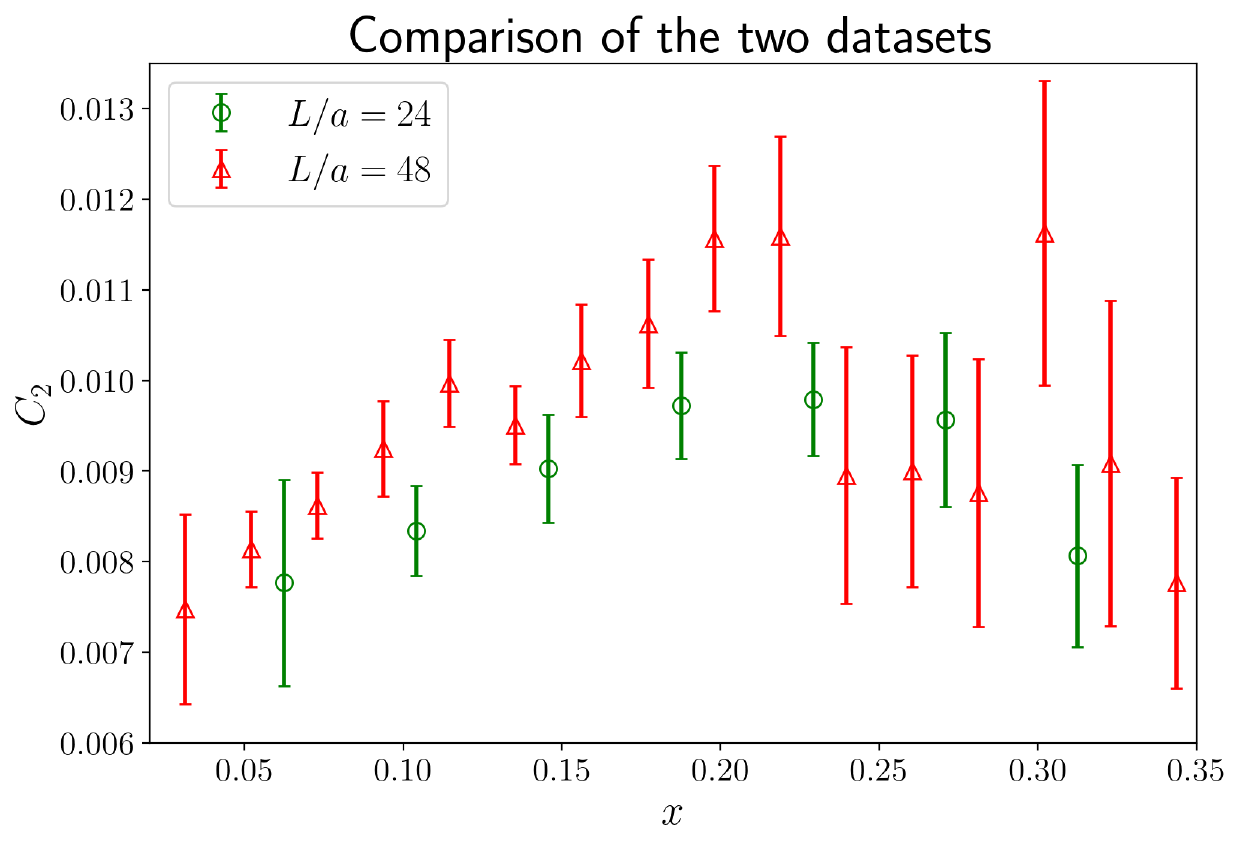}
\caption{}
\label{subfig:comparison_c-fun_3d}
\end{subfigure}
\caption{Comparison between our data and theoretical curves. (a): R\'enyi entropy for $\frac{L}{a}=48$ calculated by numerical integration of $\Delta S_2$; inset: R\'enyi entropy for $\frac{L}{a}=24$. (b) and (c): entropic c-function. (d): comparison between data for $\frac{L}{a}=24$ and for $\frac{L}{a}=48$ at small values of $x$. For $\frac{L}{a}=24$ $N$ and $n_t$ span from $2.5\times 10^4$ and $50$ for $x$ close to $0$ and $1$, to $10^5$ and $100$ for $x$ close to $\frac{1}{2}$, whereas for  $\frac{L}{a}=48$ $N$ and $n_t$ span from $4\times 10^4$ and $50$ for $x$ close to $0$ and $1$, to $2\times 10^5$ and $200$ for $x$ close to $\frac{1}{2}$.}
\label{fig:fit_c-fun_3d}
\end{figure}\noindent
%
%
Since no analytical solution is known for the entropic c-function of the three-dimensional Ising model, it is interesting to compare our data to analytical results that were derived in other simpler models. It is also instructive to study the second R\'enyi entropy, obtained by numerical integration of the contributions $\Delta S_2$ computed by means of our algorithm, in order to have a more direct comparison with results existing in the literature.

The fit functions we used to study the R\'enyi entropy are
\begin{align}
g_{\mbox{\tiny{2D}}}(x;c,k) &= c\log(\sin(\pi x)) + k, \label{fit_entropy_2d_analogue_3d} \\
g_{\mbox{\tiny{RVB}}}(x;c,k) &= -2 c \log\left\{ \frac{\eta(\tau)^2}{\theta_3(2\tau)\theta_3(\tau/2)} \frac{\theta_3(2x\tau)\theta_3(2(1-x)\tau)}{\eta(2x\tau)\eta(2(1-x)\tau)}\right\} + k, \label{fit_entropy_RVB_3d} \\
g_{\mbox{\tiny{AdS}}}(x;c,k) &= c \chi(x)^{-\frac{1}{3}}\left\{ \int_0^1\frac{\dd y}{y^2}\left( \frac{1}{\sqrt{P(\chi(x),y)}} - 1 \right) - 1 \right\} + k, \label{fit_entropy_AdS_3d}
\end{align}
where in eq.~(\ref{fit_entropy_RVB_3d}) $\tau$ is the modular parameter of the torus and in our work $\tau = i$ since $L_y = L_x$, while in the last function $P(\chi,y)=1-\chi y^3 - (1-\chi)y^4$ and
\begin{align}
x(\chi) = \frac{3}{2\pi}\chi^{\frac{1}{3}}(1-\chi)^{\frac{1}{2}}\int_0^1\frac{\dd y y^2}{(1-\chi y^3)}\frac{1}{\sqrt{P(\chi,y)}}.
\end{align}
According to the definition~(\ref{symmetrised_entropic_c-function}), the fit functions for the entropic c-function are, respectively
\begin{align}
f_{\mbox{\tiny{2D}}}(x;c) &= \frac{c}{2\pi}\sin(\pi x)\cos(\pi x), \label{fit_c-fun_2d_analogue_3d} \\
f_{\mbox{\tiny{RVB}}}(x;c) &= \frac{\sin(\pi x)^2}{2\pi^2}\frac{\dd }{\dd x}g_{\mbox{\tiny{RVB}}}(x;c,k), \label{fit_c-fun_RVB_3d} \\
f_{\mbox{\tiny{AdS}}}(x;c) &= \frac{\sin(\pi x)^2}{2\pi^2}\frac{\dd }{\dd x}g_{\mbox{\tiny{AdS}}}(x;c,k). \label{fit_c-fun_AdS_3d}
\end{align}
The functions in eq.~(\ref{fit_entropy_2d_analogue_3d}) and in eq.~(\ref{fit_c-fun_2d_analogue_3d}) are the same functions of the two dimensional case, up to a $\sin(\pi x)$ term in the entropic c-function (which arises from our definition), while the function in eq.~(\ref{fit_entropy_RVB_3d}) was found by studying the ground state of resonance-valence-bond (RVB) dimers~\cite{Stephan:2013eig}, a model in the quantum Lifshitz universality class~\cite{Ardonne:2003wa} for which the $\frac{Z_n}{Z^n}$ ratio appearing in the R\'enyi entropies reduces to a ratio of partition functions of a free $D=2$ CFT. Inglis and Melko~\cite{Inglis:2013eaa} compared the functions in  eq.~(\ref{fit_entropy_2d_analogue_3d}) and eq.~(\ref{fit_entropy_RVB_3d}) to the R\'enyi entropy of the transverse-field Ising model in $(2+1)$ dimensions, calculated by means of a quantum Monte Carlo algorithm, finding good agreement particularly in the latter case. Finally, Chen \textit{et al.}~\cite{Chen:2014zea} used the Ryu-Takayanagi formula~\cite{Ryu:2006bv} to determine the correction to the area law for theories that admit a holographic dual, finding the functional form in eq.~(\ref{fit_entropy_AdS_3d}). Then, they compared this function, as well as the RVB result, to the entanglement entropy calculated numerically in two different free fermionic theories, finding good agreement in both cases.

Note that the derivative of both $g_{\mbox{\tiny{RVB}}}$ and $g_{\mbox{\tiny{AdS}}}$ diverges as $x^{-2}$ for $x\to 0$, hence the entropic c-function converges to a non-zero constant. The 2D function has a different behavior, since $\frac{\dd}{\dd x}g_{\mbox{\tiny{2D}}}$ diverges as $x^{-1}$, hence the entropic c-function tends to zero for $x\to 0$. In recent works it was argued that in three-dimensional systems on a torus a $x^{-2}$ divergence is expected~\cite{Chen:2016kjp, Kulchytskyy:2019hft}, hence eq.~(\ref{fit_entropy_2d_analogue_3d}) and eq.~(\ref{fit_c-fun_2d_analogue_3d}) are expected to provide a worse modelling of the data.

In table~\ref{tab:fit_c-fun_3d} we collect the results of the fits shown in fig.~\ref{fig:fit_c-fun_3d}.
\begin{table}[H]
\begin{center}
\begin{tabular}{|c|c|c|c|}
\hline
 & \multicolumn{3}{|c|}{$g_{\mbox{\tiny{2D}}}$}\\
\hline
$L$  & $\frac{\chi^2}{\nu}$ & $c$ & $k$ \\
\hline \hline
$24$ & $1.07$ & $0.139(2)$ & $0.3612(16)$ \\
\hline
$48$ & $19.61$ & $0.198(7)$ & $0.750(8)$ \\
\hline
\end{tabular}
\begin{tabular}{|c|c|c|c|}
\hline
 & \multicolumn{3}{|c|}{$g_{\mbox{\tiny{RVB}}}$}\\
\hline
$L$ & $\frac{\chi^2}{\nu}$ & $c$ & $k$ \\
\hline \hline
$24$ & $4.66$ & $0.078(2)$ & $0.331(3)$ \\
\hline
$48$ & $8.89$ & $0.0713(16)$ & $0.692(4)$ \\
\hline
\end{tabular}

\begin{tabular}{|c|c|c|c|}
\hline
 & \multicolumn{3}{|c|}{$g_{\mbox{\tiny{AdS}}}$} \\
\hline
$L$  & $\frac{\chi^2}{\nu}$ & $c$ & $k$ \\
\hline \hline
$24$ & $1.55$ & $0.115(2)$ & $0.400(3)$ \\
\hline
$48$ & $5.32$ & $0.1092(19)$ & $0.759(4)$ \\
\hline
\end{tabular}

\begin{tabular}{|c|c|c|}
\hline
 & \multicolumn{2}{|c|}{$f_{\mbox{\tiny{2D}}}$}\\
\hline
$L$ & $\frac{\chi^2}{\nu}$ & $c$ \\
\hline \hline
$24$ & $3.42$ & $0.135(5)$ \\
\hline
$48$ & $12.08$ & $0.167(10)$ \\
\hline
\end{tabular}
\begin{tabular}{|c|c|c|}
\hline
 & \multicolumn{2}{|c|}{$f_{\mbox{\tiny{RVB}}}$} \\
\hline
$L$ & $\frac{\chi^2}{\nu}$ & $c$ \\
\hline \hline
$24$ & $12.6$ & $0.079(5)$ \\
\hline
$48$ & $8.23$ & $0.074(4)$ \\
\hline
\end{tabular}
\begin{tabular}{|c|c|c|}
\hline
 & \multicolumn{2}{|c|}{$f_{\mbox{\tiny{AdS}}}$} \\
\hline
$L$ & $\frac{\chi^2}{\nu}$ & $c$ \\
\hline \hline
$24$ & $3.59$ & $0.120(4)$ \\
\hline
$48$ & $4.23$ & $0.115(4)$ \\
\hline
\end{tabular}
\end{center}
\caption{First three tables: best-fit results for the R\'enyi entropies (fig.~\ref{subfig:fit_entropy_3d}). Last three tables: best fit results for the entropic c-functions (fig.~\ref{subfig:fit_c-fun_24_3d} and fig.~\ref{subfig:fit_c-fun_48_3d}).}
\label{tab:fit_c-fun_3d}
\end{table}\noindent
For $L=24$ the function yielding the best approximation of our data is the two-dimensional one, seemingly in contradiction with the expectations from ref.~\cite{Inglis:2013eaa} and with the previous discussion; however this can be interpreted as a non-universal, finite-size effect: for $L=48$ the 2D functions fare more poorly, while the AdS function seems to be the one that provides the best description of the data.

It is interesting to investigate the behavior of the coefficient $c$ when the lattice size is varied. As observed in ref.~\cite{Inglis:2013eaa}, in the 2D case it has a clear dependence on $L$ and it does not seem to converge to a constant for large $L$, while the same constant in the case of the RVB function has a less pronounced dependence on $L$. Our results exhibit the same behavior and, additionally, also reveal that the coefficient $c$ in the AdS function has a milder dependence on $L$ compared to the 2D function: these quantities are thus expected to encode universal information.

%
%
\begin{figure}[t]
\centering
\includegraphics[width=.5\textwidth]{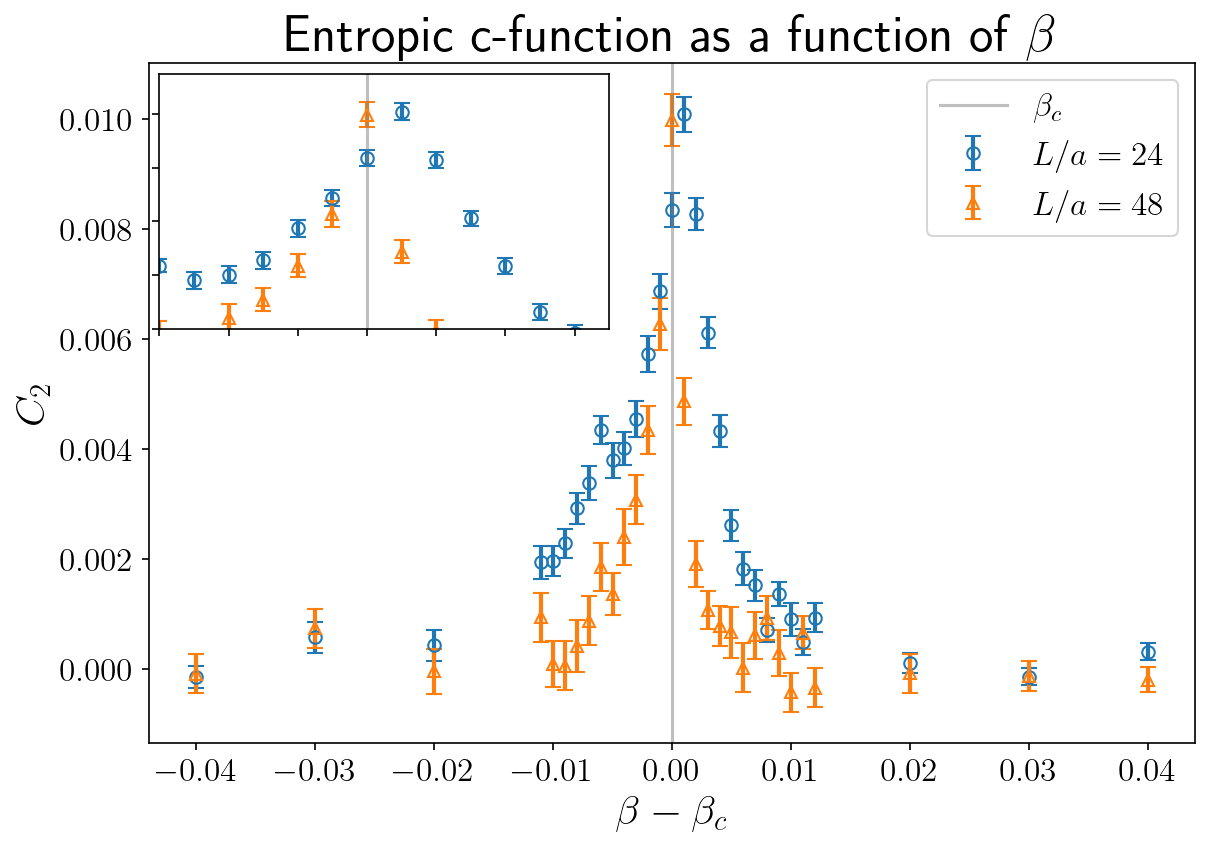}
\caption{$C_2$ as a function of $\beta$ for $L = 24$, $48$ and $l=3$, $6$ respectively. $N$ and $n_t$ are equal to $2.5\times 10^4$ $(8\times 10^4)$ and $50$ $(75)$ for $L=24$ $(L=48)$. Inset: zoom on the critical region.}
\label{fig:beta_3d}
\end{figure}\noindent
%
%
In a recent work, Kulchytskyy \textit{et al.}~\cite{Kulchytskyy:2019hft} studied a constant that can be extracted from the R\'enyi entropy of a torus in the universality class of the three-dimensional Ising model. Under the assumption
\begin{align}
S_n(x\to 0, L) = a_n \frac{L}{a} - \frac{k_n}{x} + \dots
\label{Melko_scaling_form_3D_entropy}
\end{align}
they estimated the coefficient $k_2 = 0.0174(5)$; in terms of the entropic c-function, this coefficient can be evaluated as $k_2 = 2 C_2(x\to 0)$. It should be noted that the result of ref.~\cite{Kulchytskyy:2019hft} was obtained in the $L\to \infty$ limit, while in our work we did not perform such extrapolation, hence we do not have control on finite-size corrections in our data. With this \textit{caveat} in mind, we can compare our results for $2 C_2(x)$ at small values of $x$ to $k_2$.
\begin{table}[H]
\begin{center}
\begin{tabular}{|c|c|}
\hline
$x$ & $2C_2$, $L=24$ \\
\hline
$0.0625$ & $0.015(2)$ \\
\hline
$0.104167$ & $0.0167(10)$ \\
\hline
$0.145833$ & $0.0181(12)$ \\
\hline
$0.1875$ & $0.0195(12)$ \\
\hline
\end{tabular}
\begin{tabular}{|c|c|}
\hline
$x$ & $2C_2$, $L=48$ \\
\hline
$0.0520833$ & $0.0163(8)$ \\
\hline
$0.0729167$ & $0.0172(8)$ \\
\hline
$0.09375$ & $0.0185(11)$ \\
\hline
$0.114583$ & $0.0200(9)$ \\
\hline
\end{tabular}
\end{center}
\caption{$2C_2$ for small values of $x$.}
\label{tab:2C2_small_x}
\end{table}\noindent
From table~\ref{tab:2C2_small_x} we can see that our data are close to the estimate of ref.~\cite{Kulchytskyy:2019hft}.

\subsubsection*{Variation with $\beta$}
Finally, let us discuss the behavior of the entropic c-function away from the criticality. As shown in fig.~\ref{fig:beta_3d}, our data suggest that also in three dimensions $C_2$ behaves as a Zamolodchikov c-function, i.e., it is monotonically decreasing when we move away from the critical point. The result is non-trivial: while there exist proofs of the monotonicity of entropic c-functions in three dimensions~\cite{Casini:2012ei}, they are not shape-independent, and typically assume that $A$ is a circle, not a slab as in our setup.

\section{Conclusions}
\label{sec:conclusions}

In this work we studied a new numerical algorithm for lattice calculations of the entropic c-function, a quantity related to the derivative of the R\'enyi entropy, which is expected to encode universal information on the ground state of a quantum many-body system and to provide a suitable measure of the number of effective degrees of freedom of a theory. Our primary goal consisted in showing that non-equilibrium Monte Carlo simulations based on Jarzynski's theorem~\cite{Jarzynski:1996oqb, Jarzynski:1997ef} provide a reliable and efficient tool to estimate such quantity numerically.

As a first benchmark of our algorithm, we studied the Ising model in two dimensions, for which exact analytical predictions for various quantities are known, especially at the critical point, where the entropic c-function takes the remarkably simple form expressed by eq.~(\ref{entropic_c-function_on_the_cylinder_2D}). Our results perfectly match the theory once finite-size corrections are taken into account. By exploiting GPU parallelization and running on supercomputing clusters, we obtained very precise results in a small amount of time: our data for the simulations of a $L = 128$ lattice, the largest size that we studied in $D=2$ (see fig.~\ref{fig:fit_c-fun_2d} and fig.~\ref{fig:beta_2d}), required approximately $750$ hours on the CINECA Marconi100 accelerated cluster, based on IBM Power9 architecture and Volta NVIDIA GPUs.

Albeit in two dimensions Monte Carlo methods for the calculation of entanglement entropy and related quantities are outperformed by other techniques (such as the density-matrix renormalization group), it is in $D>2$ that Monte Carlo simulations are expected to be a highly competitive tool to investigate the entanglement content of strongly correlated systems.

For this reason, we studied the entropic c-function of the $D=3$ Ising model. In this case, no analytical solution is known, and only a few numerical studies are present in literature~\cite{Inglis:2013eaa, Kulchytskyy:2019hft}. Our analysis matches results of past studies. In addition, we found that the  expression in eq.~(\ref{fit_entropy_AdS_3d}), that was derived in ref.~\cite{Chen:2014zea} using holographic methods, is, among the functions that we tested, the one that provides the best approximation of the entropic c-function of the Ising universality class.

As previously stressed, in this work we did not make any extrapolation to the infinite-volume limit, however our results for large values of the lattice size are close to the ones obtained from precise studies in the thermodynamic limit~\cite{Kulchytskyy:2019hft}, therefore we expect that our algorithm can be a valuable tool for precision studies of universal, subleading corrections to the entanglement entropy in $D \geq 3 $ quantum systems. For the three-dimensional Ising model, the total amount of computing time to obtain our results on the Marconi100 machine for $L=24$ was about $270$ hours, while for $L=48$ it was about $620$ hours.

Our work opens the path to a number of further studies. In particular, it would be interesting to apply our algorithm to compute the entropic c-function in gauge theories, whose study has been severely limited by the high numerical costs that traditional methods require~\cite{Buividovich:2008gq, Buividovich:2008kq, Buividovich:2008yv, Nakagawa:2009jk, Nakagawa:2010kjk, Itou:2015cyu, Rabenstein:2018bri} and by the ambiguity of the definition itself of the entanglement entropy in the presence of a local symmetry (see, e.g., refs.~\cite{Aoki:2015bsa, Pretko:2018nsz}). In addition, it would be interesting to compare the high-precision numerical results that can be obtained using our algorithm with the analytical predictions derived from the analysis of conformal defects~\cite{Bianchi:2015liz}. Finally, it would also be interesting to extend our method to other entanglement measures, like entanglement negativity~\cite{Calabrese:2012ew, Calabrese:2012nk, Alba:2013mg} and/or other recent proposals~\cite{Yin:2022toc}. We leave these directions of research for future works.

\subsection*{Acknowledgements}
This work has been supported by the Spoke 1 ``FutureHPC \& BigData'' of the Italian Research Center on High-Performance Computing, Big Data and Quantum Computing (ICSC) funded by MUR Missione 4 Componente 2 Investimento 1.4: Potenziamento strutture di ricerca e creazione di ``campioni nazionali di R{\&}S (M4C2-19)'' -- Next Generation EU (NGEU). The numerical simulations were run on machines of the Consorzio Interuniversitario per il
Calcolo Automatico dell'Italia Nord Orientale (CINECA). We acknowledge support from the SFT Scientific Initiative of the Italian Nuclear Physics Institute (INFN).

\appendix

\section{Jarzynski's theorem}
\label{app:jarzynski_s_theorem}
\renewcommand{\theequation}{A.\arabic{equation}}
\setcounter{equation}{0}

Jarzynski's theorem is a result of non-equilibrium statistical mechanics that allows one to express differences of free energies, or equivalently ratios of partition functions, to averages over out-of-equilibrium samples~\cite{Jarzynski:1996oqb, Jarzynski:1997ef}. Suppose we want to calculate the free-energy difference between two statistical systems described respectively by two Hamiltonians $\Hi$ and $\Hf$, that can be continuously connected to each other by tuning a parameter $\lambda$, such that $H_{\lambda = 0} = \Hi$ and $H_{\lambda = 1} = \Hf$. The system, initially at thermal equilibrium, is then driven out of equilibrium by letting the parameter $\lambda$ evolve from $0$ to $1$ at a finite rate, in an interval of time $\tf - \ti$, following a trajectory in the phase space. Along a given trajectory, we can define the work performed on the system as
\begin{align}
\mathcal{W} = \int_{\ti}^{\tf}\dd t \dot{\lambda}\frac{\partial H}{\partial \lambda}
\label{off_equilibrium_work}
\end{align}
Jarzynski's theorem then states that the exponential average of the work (in units of the temperature) that is done on the system during its off-equilibrium evolution is equal to the ratio between the equilibrium partition functions for the system at $\lambda=0$ and $\lambda=1$
\begin{align}
\left\langle \exp\left(-W\right) \right\rangle = \frac{\Zf}{\Zi},
\label{Jarzynski_theorem}
\end{align}
where we introduced $W=\frac{\mathcal{W}}{T}$. The average is calculated over the ensemble of trajectories that the system can follow given the protocol $\lambda(t)$ which determines its evolution, that is fixed for all the trajectories. It has to be noted, however, that the result in eq.~(\ref{Jarzynski_theorem}) does not depend on the choice of the protocol that, for this reason, is arbitrary.

Equation~(\ref{Jarzynski_theorem}) can be proven as follows. We assume that the system under consideration is described by a set of degrees of freedom that we collectively denote as $\phi$, and define
\begin{align}
\pi_\lambda[\phi] = \frac{1}{Z_\lambda}e^{-H_\lambda[\phi]}
\label{equilibrium_distribution}
\end{align}
as the normalized equilibrium distribution at fixed $\lambda$. Furthermore, let $P_\lambda[\phi,\phi']$ be the transition probability at fixed $\lambda$, with a normalization
\begin{align}
\sum_{\phi'}P_\lambda[\phi,\phi'] = 1.
\label{normalization_of_transition_probability}
\end{align}
For all $\phi$, $\phi'$, the detailed-balance condition is satisfied
\begin{align}
\pi_\lambda[\phi] P_\lambda[\phi,\phi'] = \pi_\lambda[\phi'] P_\lambda[\phi',\phi].
\label{detailed_balance}
\end{align}
We discretize the time interval $[\ti, \tf]$ in $N$ subintervals of equal size, such that $\epsilon = \frac{\tf - \ti}{N}$ and $t_k = \ti + k\epsilon$, with $k$ ranging from $0$ to $N$. Furthermore, let $\phi_k$ denote the configuration sampled at time $t_k$, when the parameter $\lambda$ takes the value $\lambda_k = \frac{k}{N}$ (without loss of generality we consider a linear evolution of the parameter). The work done on the system in one Monte Carlo step is given by
\begin{align}
\delta \mathcal{W}_{k} = H_{\lambda_{k+1}}[\phi_k] - H_{\lambda_k}[\phi_k],
\label{work_in_one_MC_step}
\end{align}
hence the exponential of the work (in units of $T$) along a given trajectory reads
\begin{align}
\exp(-W) = \prod_{k=0}^{N-1}\exp\left(\frac{-H_{\lambda_{k+1}}[\phi_k] + H_{\lambda_k}[\phi_k]}{T}\right) = \prod_{k=0}^{N-1}\frac{Z_{\lambda_{k+1}}\pi_{\lambda_{k+1}}[\phi_k]}{Z_{\lambda_k}\pi_{\lambda_k}[\phi_k]},
\end{align}
where in the second equality we used the definition in  eq.~(\ref{equilibrium_distribution}). To prove Jarzynski's theorem, the previous quantity has to be averaged over the statistical distribution of out-of-equilibrium trajectories, which can be built iteratively: the initial configuration is sampled from an equilibrium distribution with probability $\pi_{\lambda_0}[\phi_0]$, while the following ones are obtained by updating the system through of a sequence of transition probabilities $\pi_{\lambda_0}[\phi_0]P_{\lambda_1}[\phi_0,\phi_1] \dots P_{\lambda_{k}}[\phi_{k-1},\phi_k]$. It follows that the exponential average of $W$ can be expressed as
\begin{align}
\left\langle \exp\left( -W \right) \right\rangle = \sum_{\phi_0}\sum_{\phi_1}\dots\sum_{\phi_{N}}\pi_{\lambda_0}[\phi_0]\prod_{k=0}^{N-1} \frac{Z_{\lambda_{k+1}}\pi_{\lambda_{k+1}}[\phi_k] P_{\lambda_{k+1}}[\phi_k,\phi_{k+1}]}{Z_{\lambda_k}\pi_{\lambda_k}[\phi_k]}.
\label{discretized_average_exp_work}
\end{align}
As a side remark, note that the discretization introduces an asymmetry
in the time evolution, since the transition probability from time $t_k$ to time $t_{k+1}$ is calculated with respect to the parameter at time $t_{k+1}$.

The partition functions appearing on the right-hand side of eq.~(\ref{discretized_average_exp_work}) cancel against each other, except for the ones corresponding to the initial and the final configurations, leading to a multiplicative factor $\frac{\Zf}{\Zi}$ in front of the sums. Then, using the detailed-balance condition~(\ref{detailed_balance}) in the numerator of eq.~(\ref{discretized_average_exp_work}), one obtains
\begin{align}
\left\langle \exp\left( -W \right) \right\rangle = \frac{\Zf}{\Zi}\sum_{\phi_0}\sum_{\phi_1}\dots\sum_{\phi_{N}} \pi_{\lambda_0}[\phi_0]\prod_{k=0}^{N-1} \frac{\pi_{\lambda_{k+1}}[\phi_{k+1}] P_{\lambda_{k+1}}[\phi_{k+1},\phi_{k}]}{\pi_{\lambda_k}[\phi_k]}.
\end{align}
In this way all $\pi_\lambda$ cancel against each other, except for $\pi_{\lambda_N}[\phi_N]$, and one is left with
\begin{align}
\left\langle \exp\left( -W \right) \right\rangle = \frac{\Zf}{\Zi}\sum_{\phi_0}\sum_{\phi_1}\dots\sum_{\phi_{N}} \pi_{\lambda_N}[\phi_N]\prod_{k=0}^{N-1}P_{\lambda_{k+1}}[\phi_{k+1},\phi_k].
\end{align}
In the latter expression the term $\phi_0$ appears only in the transition probability $P_{\lambda_1}[\phi_1,\phi_0]$, hence one can use the normalization~(\ref{normalization_of_transition_probability}) to carry out the sum over $\phi_0$. The same reasoning applies for all the sums, except for the last one, $\sum_{\phi_N}$. Using the normalization of the probability distribution $\sum_{\phi_N}\pi_{\lambda_N}[\phi_N]=1$, one finally obtains
\begin{align}
\left\langle \exp\left( -W \right) \right\rangle = \frac{\Zf}{\Zi},
\end{align}
which proves eq.~(\ref{Jarzynski_theorem}). This result is crucial for our work and deserves some comments.

As previously stated, the theorem does not depend on the specific protocol $\lambda(t)$ that is used to drive the system out of equilibrium, nor on the rate $\dot{\lambda}$ at which the system evolves (in this case $\dot{\lambda} = \frac{1}{N}$). This is true for a sampling with infinite statistics, however in the case of Monte~Carlo simulations with finite statistics, this statement needs some care; in particular, it is useful to investigate two well known limiting cases of eq.~(\ref{Jarzynski_theorem}), namely $N=1$ and $N\to\infty$, in order to better understand what happens between these two limits.

Firstly, consider the $N = 1$ case. In this limit, the parameter $\lambda$ is switched directly from $\lambdai$ to $\lambdaf$ and, according to the derivation above, no off-equilibrium evolution is needed at all: the total work expressed by eq.~(\ref{work_in_one_MC_step}) is simply given by the difference between initial and final Hamiltonians, both evaluated on the initial configuration, sampled from a distribution at equilibrium. In this case, Jarzynski's theorem reduces to reweighting~\cite{Ferrenberg:1988yz}. Although reweighting is a method that in principle is exact, it is limited by large uncertainties that arise when the probability distributions $\pi_{\lambdai}$ and $\pi_{\lambdaf}$ are poorly overlapping. When this happens, the average $\langle \exp(-W) \rangle$ is dominated by very rare configurations in the tail of the initial distribution, so that (sometimes prohibitively) large statistics is needed to properly sample the target distribution.

The opposite limit is when $N\to\infty$, i.e., when the evolution is infinitely slow and the system remains in thermal equilibrium at all steps; note that two sources of asymmetry in the time evolution of the system disappear in this case: firstly, not only the initial one, but also the configurations at all intermediate steps and the final one are in thermal equilibrium; secondly, the probability distributions at time $t_k$ and $t_{k+1}$ are essentially overlapping, meaning that the aforementioned asymmetry, due to the fact that the transition probability $P_{\lambda_{k+1}}[\phi_k,\phi_{k+1}]$ depends on $\lambda_{k+1}$, is expected to vanish.

It is now more clear what happens for finite $N > 1$: in this case the system is driven out of equilibrium, leading to a statistical distribution of trajectories and fluctuations in the work calculated. The average appearing in eq.~\eqref{Jarzynski_theorem} is dominated by rare configurations, however, compared to the reweighting technique, the ensuing potential overlap problem can be mitigated by increasing $N$, which makes the distributions $\pi_{\lambda_k}$ and $\pi_{\lambda_{k+1}}$ closer to each other. Note that a simulation algorithm based on Jarzynski's theorem provides a natural way to check if the results are biased by poor sampling, simply by checking whether the results obtained letting the system evolve from $\lambdai$ to $\lambdaf$ (direct protocol) or vice~versa (reverse protocol) are consistent with each other (see also ref.~\cite{Jarzynski:2006re}):
\begin{align}
\left\langle \exp\left( -W \right) \right\rangle_{\mbox{\tiny{direct}}}= \left\langle \exp\left( -W \right) \right\rangle^{-1}_{\mbox{\tiny{reverse}}}
\label{equality_of_direct_reverse_protocols}
\end{align}

\section{A different protocol for the three-dimensional case}
\label{app:a_different_protocol_for_the_three-dimensional_case}
\renewcommand{\theequation}{B.\arabic{equation}}
\setcounter{equation}{0}

\begin{figure}[t]
\centering
\begin{subfigure}{.32\textwidth}
\includegraphics[width=\textwidth]{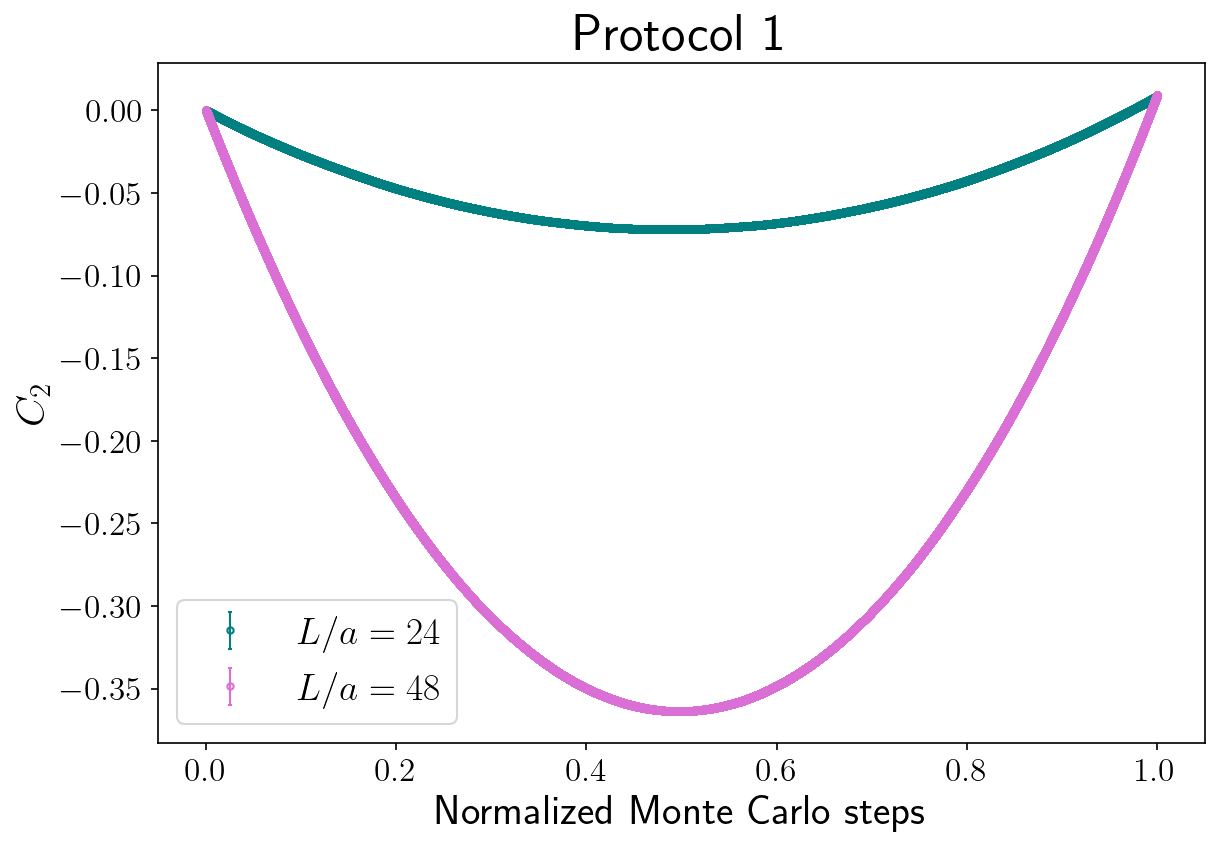}
\caption{}
\label{subfig:protocol_1}
\end{subfigure}
\begin{subfigure}{.32\textwidth}
\includegraphics[width=\textwidth]{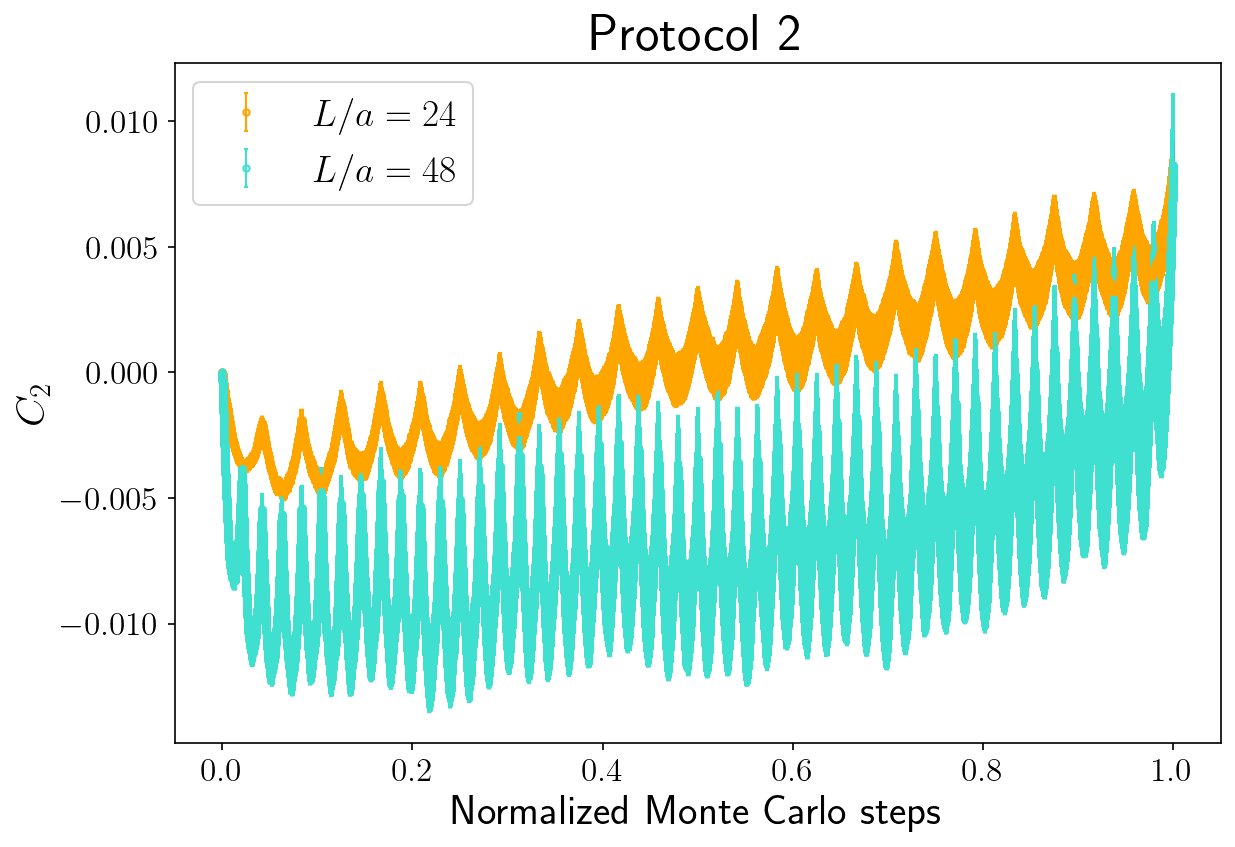}
\caption{}
\label{subfig:protocol_2}
\end{subfigure}
\begin{subfigure}{.32\textwidth}
\includegraphics[width=\textwidth]{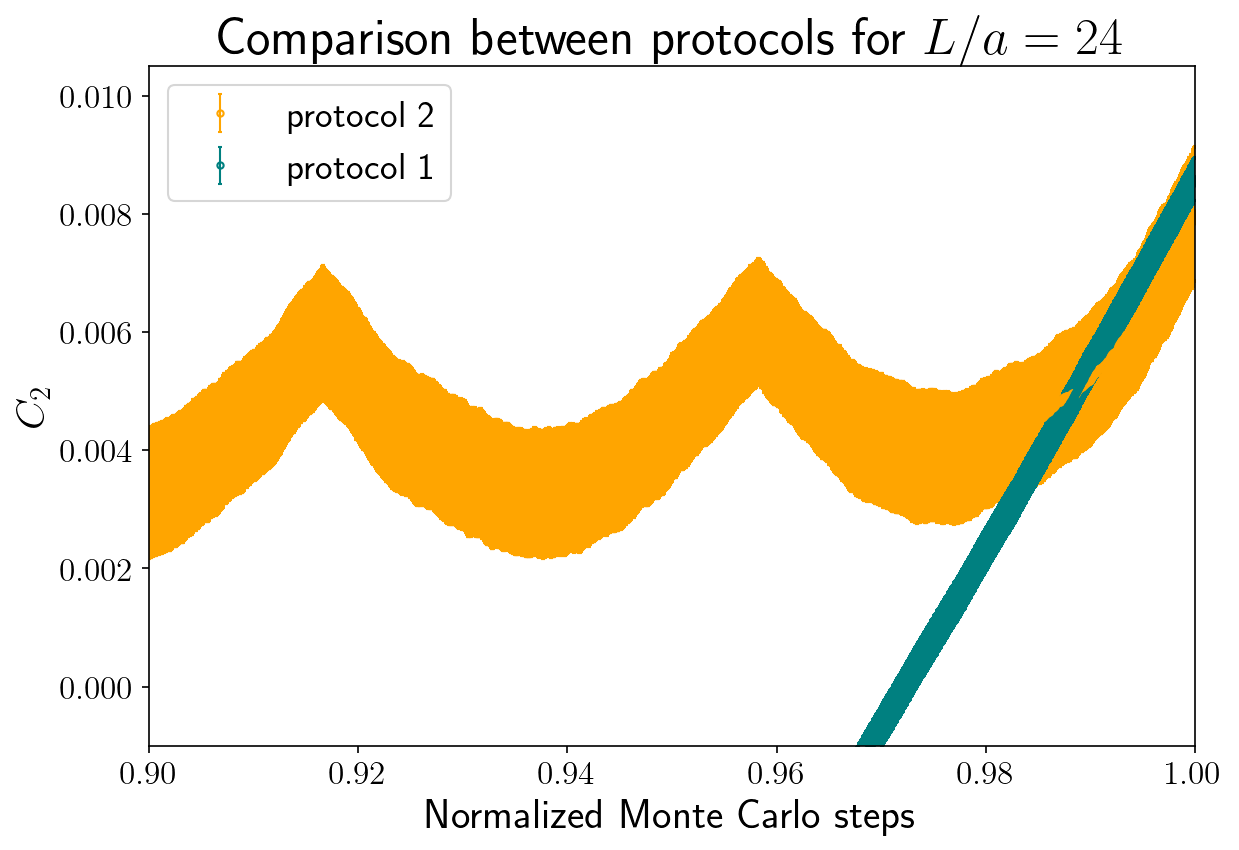}
\caption{}
\label{subfig:protocol_comparison}
\end{subfigure}
\caption{(a) and (b): off-equilibrium evolution of $C_2$ in $D=3$ for protocols $1$ and $2$, respectively. (c): comparison of the two protocols close to the end of the out-of-equilibrium evolution. For $L=24$, the parameters of the simulations were $N=2.4\times 10^4$ and $n_t=50$, whereas for $L=48$ they were $N=7.2\times 10^4$ and $n_t=75$.}
\label{fig:protocols}
\end{figure}

Lattice calculations of R\'enyi entropies and the entropic c-functions present in literature typically exploit two different techniques. The authors of refs.~\cite{Caraglio:2008pk, Alba:2011fu} used reweighting, whose limitations are discussed in section~\ref{app:jarzynski_s_theorem} of the appendix. The other technique was typically used in gauge theories~\cite{Buividovich:2008gq, Buividovich:2008kq, Buividovich:2008yv, Nakagawa:2009jk, Nakagawa:2010kjk, Itou:2015cyu, Rabenstein:2018bri} and is based on the following identity
\begin{align}
-\log\frac{\Zf}{\Zi} = -\int^1_0 \dd \lambda \frac{\partial \log Z(\lambda)}{\partial \lambda} = \int^1_0 \dd \lambda \langle \Sf - \Si \rangle_{Z(\lambda)},
\end{align}
where the interpolating partition function is defined as
\begin{align}
Z(\lambda) = \int D\phi \exp\{-(1-\lambda)\Si[\phi] - \lambda \Sf[\phi]\}.
\end{align}
It was pointed out in a recent work~\cite{Rindlisbacher:2022bhe} that this method is affected by a bad signal-to-noise ratio since, as $\lambda$ grows, the algorithm has to sample configurations separated by a large energy barrier. Motivated by this observation, we also studied the behavior of $C_2$ in $D=3$ as a function of the evolution parameter $\lambda$ during the out-of-equilibrium evolution, finding an energy barrier also in our case. As shown in fig.~\ref{subfig:protocol_1}, the barrier grows with the lattice size; in particular the ratio between the peak of the barrier and the final value of $C_2$ is about $10$ for $L=24$, while it is about $50$ for $L=48$ at $x$ close to $0$, where $C_2$ takes its maximum value. While these values of the height of the peak are not extremely large, we tried to implement a different simulation strategy, inspired by ref.~\cite{Rindlisbacher:2022bhe}, in order to avoid the barrier.

Notice that in the protocol used in section~\ref{sec:Ising_3D} for the three-dimensional Ising model (which, from now on, we will call protocol $1$) all the $4L$ couplings at the boundary between $A$ and $B$ are simultaneously varied. As previously noticed, this is not strictly necessary: indeed one can divide the couplings in subsets and then let them evolve one subset at a time. More specifically, in the alternative protocol we tested (protocol $2$) we divided the whole evolution in $L$ different subsets, in which only $4$ couplings are varied, exactly as in the two-dimensional case; the number of Monte Carlo steps in which the evolution of a subset takes place is $m = \frac{N}{L}$.

We can then compare the two protocols at fixed $N$. Using the protocol $2$, a single coupling is driven far from equilibrium more rapidly, however at a given time a smaller number of couplings is varied, compared to protocol $1$.

%
%
\noindent
%
%
Fig.~\ref{subfig:protocol_2} shows the off-equilibrium evolution of $C_2$ using protocol $2$: the energy barrier has effectively been avoided and the evolution is now characterized by small jumps. However one can see that at each jump the statistical error increases, in contrast to what happens in simulations based on the protocol $1$. This behavior can be explained as follows: using the protocol $2$, the evolution of each subset can be seen as an independent out-of-equilibrium evolution, but only the first one starts from a thermalized configuration. This implies that all subsequent configurations are further and further away from equilibrium, leading to a larger dispersion of trajectories. In particular, one can note that the statistical uncertainties obtained from simulations based on the protocol $2$ are larger than those from protocol $1$ by a factor about $3$ for $L = 24$, and about $5$ for $L=48$, see fig.~\ref{subfig:protocol_comparison}. A simple way to fix the problem is to sample the first step of the evolution from an equilibrium distribution: this is nothing but the increment trick, with the increment performed in the direction orthogonal to the entangling surface. Although an explicit study of the performances of such protocol is beyond the scopes of our work, it would be interesting to compare this algorithm with protocol $1$, as it could further improve the calculation of the entropic c-function in strongly interacting quantum field theories.

\section{Systematic uncertainty from the derivative discretization}
\label{app:systematic_uncertainty_from_the_derivative_discretization}
\renewcommand{\theequation}{C.\arabic{equation}}
\setcounter{equation}{0}

In our work the derivative of the R\'enyi entropy was estimated by means of finite differences. This introduces a systematic error: by using a mid-point approximation (and momentarily reintroducing the lattice spacing $a$ in our notations), the finite difference reads
\begin{align}
\frac{S_2(l) - S_2(l-a)}{a} = \frac{\dd}{\dd l} \eval{S_2}_{l-\frac{a}{2}} + \frac{a^2}{24}\frac{\dd^3}{\dd l^3} \eval{S_2}_{ l-\frac{l}{a}} + \order{a^4}.
\end{align}
Hence the entropic c-function can be written as
\begin{align}
C^{MC}_2(x) = C_2(x) + \frac{1}{48 L^2}\frac{\sin(\pi x)^{D-1}}{\pi^{D-1}}\frac{\dd^3}{\dd x^3}\eval{S_2}_{x} + \order{\frac{1}{L^4}},
\label{entropic_c-function_and_third_derivative}
\end{align}
where $x = \frac{l-.5}{L}$. Therefore we used the right-hand side of \eqref{entropic_c-function_and_third_derivative} to estimate the systematic uncertainty associated with the derivative discretization. The third derivative of $S_2$ was estimated \textit{a posteriori}: firstly we fitted the data, then we used the best-fit result to calculate the systematic error. The procedure was repeated until convergence of the systematic uncertainty.

In all cases that we considered in this work, the third derivative of the fit function is divergent at $x=0$ and at $x=1$, then it rapidly becomes zero. Since the divergence dominates in the $\sin(\pi x)^{D-1}$ term, the systematic error is expected to become relevant when $x$ gets close to $0$ or to $1$.

In $D=2$ we chose to estimate the systematic error using the function in eq.~\eqref{fit_function_2d}, taking into account the leading scaling correction; while the scaling function converges to a constant, the non-universal term is divergent for $x=0$ and for $x=1$, leading to a larger systematic uncertainty. To check the validity of our method, we also computed the systematic error by comparing our results to a higher-order approximation of the derivative, which is accurate to order $\order{a^4}$: this method has the advantage of being model-independent, but it requires the evaluation of a  larger number of free-energy differences; the systematic error obtained using this latter approach is nearly equal to the one calculated \textit{a posteriori} with the function in eq.~\eqref{fit_function_2d}.

In $D=3$ the systematic uncertainties evaluated using the RVB function in eq.~\eqref{fit_entropy_RVB_3d} and the AdS function in eq.~\eqref{fit_entropy_AdS_3d} are the same, since they both diverge as $x^{-2}$ for $x \to 0$, while the 2D function in eq.~\eqref{fit_entropy_2d_analogue_3d} tends to $x^{-1}$ in that limit, hence it gives a significantly smaller contribution to the systematic error. Comparison with higher-order approximations of the derivative shows that the most accurate estimate of the systematic error is obtained using the RVB and the AdS functions.

\bibliography{paper}

\end{document}